\documentclass[11pt]{article}
\usepackage{amssymb, amsmath}
\usepackage{graphicx}
\newcommand{\bm}[1]{\mbox{\boldmath $#1$}}
\begin{document}
\makeatletter
\@addtoreset{equation}{section}
\def\theequation{\thesection.\arabic{equation}}
\def\@maketitle{\newpage
 \null
 {\normalsize \tt \begin{flushright} 
  \begin{tabular}[t]{l} \@date  
  \end{tabular}
 \end{flushright}}
 \begin{center} 
 \vskip 2em
 {\LARGE \@title \par} \vskip 1.5em {\large \lineskip .5em \begin{tabular}[t]{c}\@author 
 \end{tabular}\par} 
 \end{center}
 \par
 \vskip 1.5em} 
\makeatother
\topmargin=-1cm
\oddsidemargin=1.5cm
\evensidemargin=-.0cm
\textwidth=15.5cm
\textheight=22cm
\setlength{\baselineskip}{16pt}
\title{Holographic Duality for  3D Spin-3 Gravity Coupled to  Scalar Field
}
\author{
Ryuichi~{\sc Nakayama}\thanks{nakayama@particle.sci.hokudai.ac.jp},  Kenji~{\sc Shiohara}\thanks{k-shiohara@particle.sci.hokudai.ac.jp} and Tomotaka~{\sc Suzuki}\thanks{t-suzuki@particle.sci.hokudai.ac.jp} 
       \\[1cm]
{\small
    Division of Physics, Graduate School of Science,} \\
{\small
           Hokkaido University, Sapporo 060-0810, Japan}
}
\date{
EPHOU-19-002 \\
February  2019
}
%
%
\maketitle

\begin{abstract} 
 The 3d spin-3 gravity theory is holographically dual to a 2d ${\cal W}_3$-extended CFT. In a large-c limit the symmetry algebra of the CFT reduces to $SU(1,2) \times SU(1,2)$. On the  ground of symmetry the dual bulk space-time will be given by an 8d group manifold $SU(1,2)$. Hence we need to introduce five  extra coordinates in addition to three ordinary ones.  The 3d space-time is a 3d hyper-surface $\Sigma$ embedded at constant values of the extra variables. Operators in the CFT at the boundary of $\Sigma$ are expressed in terms of  ${\cal W}$ descendants of the operators at the boundary of $\Sigma_0$,  where the extra variables vanish. In this paper it is shown that AdS/CFT correspondence for a scalar field coupled to 3d spin-3 gravity is realized in this auxiliary 8d space. A bulk-to-boundary propagator of a scalar field is found and  a generating functional of boundary two-point functions of scalar ${\cal W}$-descendant operators is obtained  by using the  classical  action for the scalar field. Classically, the scalar field must satisfy both Klein-Gordon equation and a third-order differential equation, which are related to the quadratic and cubic Casimir operators of $su(1,2)$. It is found that the coefficient function of the derivatives of the scalar field in the latter equation is  the spin-3 gauge field, when restricted to the hypersurface. An action integral in the 8d  auxiliary space for the 3d spin-3 gravity coupled to  a scalar field is presented. In general, this 8d auxiliary space is a deformation of the manifold $SU(1,2)$.   An 8d local frame is introduced and the equations of motion for the 8d connections $A_{\mu}$, $\overline{A}_{\mu}$ are solved. By restricting those solutions onto $\Sigma$, flat connections  in 3d $SL(3,\mathbb{R}) \times SL(3,\mathbb{R})$ Chern-Simons theory are obtained and new 3d black hole solutions with and without spin-3 charge are found by this method. 

\end{abstract}
\newpage
\setlength{\baselineskip}{18pt}
\section{Introduction}
\hspace*{5mm}
After the discovery of  AdS/CFT correspondence \cite{Maldacena} this subject has been studied extensively and a lot of evidences have been accumulated until now. It is, however, 
still necessary to extend the range of applicability of this correspondence.  One of those possible directions will be the 3d spin-3 gravity\cite{Blencowe}\cite{Campoleoni} coupled to a matter (scalar) field. This is a 3d version of Vasiliev's theory\cite{Vasiliev} with higher spins consistently truncated up to 3. 
 In this theory the gravity and spin-3 gauge field are described by $SL(3,\mathbb{R}) \times SL(3,\mathbb{R})$ Chern-Simons gauge theory. However,  the action integral for a scalar field which has spin-3 charge has not been found. Hence a check of duality for the  correlation functions of conformal field theory (CFT) primary operators in the boundary ${\cal W}_3$ conformal field theory (CFT) has not been carried out except for those of the currents. Therefore it is necessary to have a Lagrangian formulation of this coupled system.  Study of the formulation of such a theory will also elucidate the nature of the 3d spin-3 gravity. 

There is also another attempt to construct 2d CFTs which are dual to 3d higher-spin gravity theories.\cite{GG} These 2d theories are based on 2d ${\cal W}_N$-minimal models which are obtained in terms of cosets of the form: 
\begin{equation}
\frac{SU(N)_k \otimes SU(N)_1}{SU(N)_{k+1}}. 
\end{equation}  
This CFT has a central charge 
\begin{equation}
c_{N,k}= (N-1)\Big[1-\frac{N(N+1)}{(N+k)(N+k+1)}\Big] \leq (N-1).  \label{cNk}
\end{equation}
This model has ${\cal W}_N$ algebra as a symmetry algebra. This is a special case of an extended symmetry algebra ${\cal W}_{\infty}[\mu]$, which has all integer spins $s \geq 2$ and which can be truncated to ${\cal W}_N$ for $\mu=N$. In the case of ${\cal W}_3$ algebra $N$ must be set to 3. The corresponding bulk theory is based on so-called $hs[\mu]$ algebra \cite{PV} and a scalar field with a suitable mass can be consistently coupled to the higher-spin gravity at the level of equations of motion. 

In this paper the 3d spin-3 gravity theory \cite{Blencowe}\cite{Campoleoni} which has a spin-3 gauge field in addition to a gravity field as well as  a scalar field will be considered. 
On the boundary of AdS$_3$ space in the spin-3 gravity the CFT has an additional symmetry, ${\cal W}_3$ symmetry. This is a non-linear algebra.\cite{Zamolodchikov}\cite{BouwknegtSchoutens} 
\begin{eqnarray}
\ [ L_m, L_n] &=& (m-n) \, L_{m+n}+\frac{c}{12}m(m^2-1) \,  \delta_{m+n,0}, \nonumber \\
\ [L_m, W_n] &=& (2m-n) \, W_{m+n}, \nonumber \\
\ [ W_m, W_n] &=& \frac{c}{36}m(m^2-1)(m^2-4) \, \delta_{m+n,0}\nonumber \\  &&+\frac{1}{3}(m-n)(2m^2+2n^2-mn-8) \, L_{m+n} \nonumber \\ &&+10 \bm{\beta} (m-n)\Lambda_{m+n}, \qquad (m,n \in \mathbb{Z}) \label{nonlinerW3}
\end{eqnarray}
 Here $\Lambda_m$ is a normal-ordered operator 
\begin{equation}
\Lambda_m =\sum_{n \leq -2}L_n L_{m-n}+\sum_{n \geq -1}L_{m-n}L_n-\frac{3}{10}(m+3)(m+2) \, L_m,
\end{equation}
and $\bm{\beta}$  in the last line of (\ref{nonlinerW3}) is a constant related to the central charge $c$ as $\bm{\beta} = 16/(22+5c)$.
In the semi-classical limit $c  \rightarrow \infty$, $\bm{\beta}$  can be dropped and the sub-algebra of the wedge modes, $L_n \ (n=0, \pm 1)$ and $W_m \ (m=0, \pm 1, \pm2)$,  survives. It is a linear $su(1,2)$ algebra  and given by 
\begin{eqnarray}
\ [L_m,L_n] &=& (m-n) \, L_{m+n}, \qquad 
\ [L_m,W_n] = (2m-n) \, W_{m+n}, \nonumber \\
\ [W_m,W_n] &=& \frac{1}{3} (m-n) \, \{2m^2+2n^2-mn-8\} \, L_{m+n}.   \label{W3wedgealgebra}
\end{eqnarray}
Actually, $su(1,2)$ is one of the two real forms of $sl(3,\mathbb{C})$, distinct from $sl(3,\mathbb{R})$.\cite{Campoleoni}\footnote{Conventions for $sl(3,\mathbb{R})$ and $su(1,2)$ used in this paper are given in appendices A and B. The difference of the two is the sign of the right hand side of the commutators $[W_m,W_n]$.} By combining the left and right sectors, the boundary field theory has global $su(1,2) \times su(1,2)$ symmetry. Then by the principle of holography the bulk spin-3 gravity theory is also expected to have $su(1,2) \times su(1,2)$ asymptotic symmetry.\footnote{In \cite{NS} it was assumed that the symmetry of the spin-3 gravity is  $sl(3,\mathbb{R}) \times sl(3,\mathbb{R}))$. The symmetry of the bulk space-time is, however,  $su(1,2) \times su(1,2)$, while $sl(3,\mathbb{R}) \times sl(3,\mathbb{R})$ is a symmetry of the local frame. Translation from $sl(3,\mathbb{R})$ to $su(1,2)$ is simply carried out by analytic continuation of some variables. See footnote 7.}

It was shown in  \cite{Campoleoni} that the 3d higher-spin gravity theory  can be  formulated as a  Chern-Simons gauge theory. 
\begin{eqnarray}
S &=& S_{CS}[A]-S_{CS}[\overline{A}],  \\
S_{CS}[A] &=& \frac{k}{4\pi} \int \text{tr} \Big( A \wedge dA +\frac{2}{3} \, A \wedge A\wedge A\Big), \label{CS}
\end{eqnarray}
where $k=\ell_{\text{AdS}}/4G$ and $\ell_{\text{AdS}}$ is the AdS length\footnote{In this paper $\ell_{\text{AdS}}$ will be sometimes set to 1. }.  
The gauge group acts on the local frame fields, $A=\omega+e$ and $\overline{A}=\omega-e$, and in the case of the spin-N gravity the gauge group is $SL(N,\mathbb{R}) \times SL(N,\mathbb{R})$. The gauge connections must satisfy suitable conditions  on the boundary in order for the  boundary value problem to be well-posed. The usually adopted boundary condition is  $A_-=\overline{A}_+=0$.  In \cite{Kraus}\cite{BH} an asymptotically AdS$_3$ solution which shows UV/IR interpolating behavior were found and a  black hole solution with spin-3 charge was also obtained. In these solutions  all components $A_{\pm}$ and $\overline{A}_{\pm}$ do not vanish and the boundary conditions are imposed on the components of connections $A_{\pm}=\sum_{a=1}^8 A_{\pm}^at_a$ separately, where $t_a$ is a generator of $sl(3,\mathbb{R})$, and the ${\cal W}_3$ algebra of the higher-spin currents in the CFT is realized as Ward identities of the currents  in the presence of a perturbation term $\int d^2z \mu W$ in the action integral. See \cite{BC}. 

When matter fields such as scalar fields are coupled to higher-spin gravity, it is still possible to describe matter degrees of freedom  by means of Wilson lines.\cite{FKLW}\cite{Castro2}  However, a natural description of matter degrees of freedom  and their coupling to spin-3 gravity in terms of Lagrangian formalism are still missing. For matter fields which have spin-3 charges it is not possible to write down an action which is  invariant under spin-3 gauge transformations as well as diffeomorphisms. It is not possible to derive boundary conformal field theory (CFT) correlation functions from on-shell action by using the standard differentiating dictionary of holography, either. The purpose of this paper is to improve this situation. The action integral for scalar fields and spin-3 gravity fields are presented in a higher-dimensional setting, which will be explained below. It is shown that when a solution to an equation of motion for a scalar field is substituted into the scalar action, a generating functional for a two-point function of scalar operators is obtained semiclassically. The source functions work as boundary conditions of the scalar fields. 

There is another motivation for the present work. When the symmetry of the 2d CFT is ${\cal W}$-extended, all states of a scalar field  must be reconstructed\cite{BDHM}\cite{HKL} in the bulk from ${\cal W}$-descendants of a scalar primary state in the boundary CFT.  In this paper we  will introduce an 8d auxiliary space dual to ${\cal W}_3$ CFT by following  our previous paper\cite{NS}. Usually in 2d (Euclidean) CFT without ${\cal W}_3$ symmetry, a representation of global Virasoro algebra is constructed as follows. In a highest-weight representation a highest-weight state $|h\rangle$ which satisfies $L^h_0 \, |h\rangle=h \, |h\rangle$ and $L^h_1 \, |h\rangle=0$ is introduced. Here the generators with a superscript $h$ represent those in the hyperbolic representation\cite{GT}, which is appropriate for Lorentzian Poincar\'e coordinates. These are defined in (2.13) and (2.18) of \cite{NS}. $L^h_{\pm 1}$ and $L^h_0$ are global  Virasoro generators. Then any states in this representation are given by  linear combinations of descendants $(L^h_{-1})^n|h\rangle$ ($n=0,1,2,\ldots$). These states are combined into a single state $|\phi(x)\rangle = \exp \{ ix L^h_{-1}\} |h\rangle$ by introducing a coordinate $x$. A shift in $x$ corresponds to a translation. In the case of a global large-c ${\cal W}_3$ algebra highest-weight representation is defined by a highest-weight state$|h,q\rangle$, which satisfies  $L^h_0 |h,q\rangle =h  |h,q\rangle$, $W^h_0 |h,q\rangle =q  |h,q\rangle$, and $L^h_1\, |h,q\rangle=W^h_1 \, |h,q\rangle=W^h_2 \, |h,q\rangle=0$.\footnote{In \cite{NS} $W_3$-charge was denoted as $\mu$. In this paper it will be denoted as $q$ instead. Later $\mu$ is used for a chemical potential.} Any descendant states in this  representation are linear combinations of states of a form $(L^h_{-1})^n(W^h_{-1})^m(W^h_{-2})^{\ell} |h,q\rangle$. By introducing variables $x$, $\alpha$ and $\beta$, these states are combined into a single state $|\phi(x,\alpha,\beta)\rangle= \exp\{ ixL^h_{-1}\} \, \exp \{-\alpha W^h_{-2}\} \, \exp\{ i\beta W^h_{-1}\}\, |h,q\rangle$. By using the coefficients of the Taylor expansion of this state in the variables, $x$, $\alpha$, $\beta$, any states can be obtained. 
 Due to the left and right movers it turns out it is necessary to introduce six variables to describe states in ${\cal W}_3$ extended CFT.\footnote{ Wakimoto representation of large-c ${\cal W}_3$ algebra is expressed in terms of  similar variables, whose relation to $x$, $\alpha$, $\beta$ is not known,  and 
this representation  was  used for calculating correlation functions in ${\cal W}_3$ extended CFT in \cite{FateevRibault}.}  As for the variables in the bulk it is necessary to additionally introduce a radial coordinate $y$ ($y=0$ is the boundary), which corresponds to $L^h_0+\overline{L}^h_0$, and another coordinate $\gamma$,  corresponding to $W^h_0+\overline{W}^h_0$. Hence the `bulk space-time' holographically dual to the boundary CFT with ${\cal W}_3$ symmetry is 8 dimensional. 

In the remainder of this section we will give a review of our paper \cite{NS}. At the end of this section the content of this paper will be presented. In \cite{NS} we constructed a state in the boundary ${\cal W}_3$ CFT which represents the one of a scalar field put at one point inside the bulk. In the case of ordinary AdS$_3$/CFT$_2$ without spin-3 gauge field, such a state $|\Phi(y=1,x^+=0,x^-=0)\rangle$  at the center of AdS$_3$ (in the Poincar\'e coordinates) must satisfy $sl(2,\mathbb{R})$ conditions.\cite{Miyaji}\cite{GT}\cite{V}\cite{NO} By the action of  $\exp \{ix^+L^h_{-1}\}$, $\exp \{ix^- \overline{L}^h_{-1}\}$ and $y^{L^h_0+\overline{L}^h_0}$ a state at any point in the bulk is obtained.\footnote{Light-cone coordinates are defined by $x^{\pm}=t \pm x$. In the case of black holes in sec.4 $x^{\pm}=t \pm \phi$.  } In the case of spin-3 gravity, a state of a scalar field $|\Phi(x^+,\alpha^+,\beta^+,x^-,\alpha^-,\beta^-,y,\gamma) \rangle$ must satisfy $su(1,2)$ conditions.\footnote{In \cite{NS} we solved an $sl(3,\mathbb{R})$ conditions for the state in the bulk. To convert the results of \cite{NS} to those appropriate for $su(1,2)$, it is necessary to make substitutions, $\alpha^{\pm} \rightarrow i\alpha^{\pm}$, $\beta^{\pm} \rightarrow i\beta^{\pm}$, $\gamma \rightarrow i\gamma$, $\mu \rightarrow -i\mu$.} At the center of the bulk these conditions are written as
\begin{eqnarray}
&& (L^h_n-(-1)^n \, \overline{L}^h_{-n}) |\Phi(y=1,\text{other coord's}=0) \rangle=0 \quad (n=-1,0,1), \label{LbL} \\
&&(W^h_n-(-1)^n \, \overline{W}^h_{-n})|\Phi (y=1,\text{other coord's}=0)\rangle=0 \quad (n=-2, \ldots, 2).  \label{WbW}
\end{eqnarray} 
In \cite{NS} the state of a scalar field in the boundary ${\cal W}_3$ CFT was explicitly constructed, and although this state is a formal integral expression, the existence of such a  state was established. Then by using exponentials of ${\cal W}_3$  generators the state for a scalar field at any point in the bulk was also obtained. During this work  an infinite-dimensional representation of ${\cal W}_3$ generators in the bulk in terms of differential operators were also obtained. This is presented in Appendix B, because some variables are redefined compared to those in \cite{NS} by analytic continuation.  

From the structure of the local state in the bulk it was also found that 
the scalar local state in the bulk satisfies a partial differential equation which is associated with the quadratic  Casimir operator of $su(1,2)$\footnote{See Appendix A and B for conventions.}. 
\begin{eqnarray}
\tilde{C}_2(L^h,W^h)&=&(L^h_0)^2-\frac{1}{2} \, (L^h_1 L^h_{-1}+L^h_{-1}L^h_1)-\frac{1}{8} \, (W^h_2W^h_{-2}+W^h_{-2}W^h_2)\nonumber \\
&&+\frac{1}{2} \, (W^h_1W^h_{-1}+W^h_{-1}W^h_1)-\frac{3}{4}(W^h_0)^2
\end{eqnarray}
By using\footnote{In replacing  $sl(3,\mathbb{R})$ with $su(1,2)$ the eigenvalue $\mu$ of $W_0$ must  also be replaced by $-iq$.  }
the representation (\ref{diffL})-(\ref{differentialop}) and 
\begin{equation}
\Big(\tilde{C}_2(L^h,W^h)+\tilde{C}_2(\overline{L}^h,\overline{W}^h)\Big) \, |O_{\Delta,q}\rangle=\frac{1}{2} \, \Big\{ \Delta^2-8\Delta-3q^2\Big\} \,  |O_{\Delta,q}\rangle, 
\end{equation}
where $ |O_{\Delta,q}\rangle$ is a CFT primary state on the boundary with eigenvalues $L^h_0=\bar{L}^h_0=\Delta/2$ and $W^h_0=\overline{W}^h_0=q$, 
a differential equation for a scalar field in 8d space was derived. It takes a form 
\begin{equation}
(\nabla^2-m^2) \, |\Phi \rangle=0, \label{KGsymbolical}
\end{equation}
where $\nabla^2$ is a Laplacian in a 8d space, which has a metric (\ref{8dmetric}) defined below.  $m$ is a mass of the scalar field related to $\Delta$ and $q$ by
\begin{equation}
m^2=\Delta^2-8\Delta-3q^2.   \label{massDelta}
\end{equation}
The explicit form of the equation  is presented in Appendix C. 
Conversely, $\Delta$ is given in terms of the mass and charge as
\begin{equation}
\Delta=4+\sqrt{m^2+16+3q^2}. \label{Deltamass}
\end{equation}
In \cite{NS} this equation was interpreted as a Klein-Gordon equation for a scalar field in the 8d space. Because the scalar field  transforms non-trivially under $SU(1,2) \times SU(1,2)$, its equation of motion must be formulated in 8d. The equation (\ref{KG}) coincides with Klein-Gordon equation for a scalar field in a space-time with a metric:
\begin{eqnarray}
ds_0^2 &=& g_{\mu\nu} \ dx^{\mu}dx^{\nu} \nonumber \\
&=&y^{-2} \, dy^2-y^{-4}(y^2 \cosh 2\gamma-4\beta^+\beta^-) dx^+dx^-+4y^{-4}d\alpha^+d\alpha^-         \nonumber \\
&&-y^{-2} \cosh 2\gamma d\beta^+d\beta^--4y^{-4} (\beta^+dx^+d\alpha^-+\beta^-dx^-d\alpha^+)    \nonumber \\
&& +y^{-2}\sinh 2\gamma (dx^+d\beta^-+dx^-d\beta^+)+\frac{1}{3} \, d\gamma^2, \label{8dmetric}
\end{eqnarray}
where $x^{\mu}=(x^+,x^-,y,\alpha^+,\alpha^-,\beta^+,\beta^-,\gamma)$. 
The determinant of this metric is $g=\text{det}\ g_{\mu\nu}=-\frac{1}{12} y^{-18}$. Metric (\ref{8dmetric}) is that of the group manifold SU(1,2). This is invariant under  $SU(1,2) \times SU(1,2)$ generated by  (\ref{diffL}), (\ref{differentialop}). The coordinates $(y, x^{\pm})$ are those of  SU(1,1)=AdS$_3$ which is principally embedded in SU(1,2). The additional coordinates $(\alpha^{\pm}, \beta^{\pm},\gamma)$ parametrize a coset SU(1,2)/SU(1,1).
This metric plays the role of the vacuum of gravity theory.\footnote{The metric is not sufficient to specify the vacuum. It is also necessary to identify the spin-3 field. This will be studied in sec.2.}

The ordinary pure AdS$_3$ space is the hypersurface $\Sigma_0$ embedded at $\alpha^{\pm}=\beta^{\pm}=\gamma=0$ in this 8d space. Then,  it will be natural to consider other hypersurfaces $\Sigma_{\alpha\beta\gamma}$ with constant non-vanishing $\alpha^{\pm}$, $\beta^{\pm}$ and $\gamma$ in the 8d space-time. All operators on the boundary of $\Sigma_{\alpha\beta\gamma}$ are transformed by $\exp( i\alpha^+ W^h_{-2})$, $\exp(\beta^+W^h_{-1})$ and $\exp (\frac{i}{2}\gamma W^h_0)$ compared to those on $\Sigma_0$.   An induced metric on this hypersurface $\Sigma_{\alpha\beta\gamma}$ is given by 
\begin{eqnarray}
ds_0^2\Big|_{\Sigma_{\alpha\beta\gamma}} &=& y^{-2} \, dy^2-y^{-4}(y^2 \cosh 2\gamma-4\beta^+\beta^-) dx^+dx^-. \label{per1}
\end{eqnarray}
Property of the hypersurface $\Sigma_{\alpha\beta\gamma}$ depends on the values of $\beta^{\pm}$. 
\begin{itemize}
\item For $\beta^+=\beta^- =0$ the hypersurface is AdS$_3$ with AdS length $\ell_{\text{AdS}}=1$ (in our units).
\item For  $\beta^+\beta^- <0$ the space-time  on the hypersurface is not AdS, but asymptotically AdS. 
This is a solution interpolating two vacua: one corresponding to UV CFT at $y=0$ with a AdS length $\ell_{\text{AdS}}'=\frac{1}{2} \, \ell_{\text{AdS}}=\frac{1}{2}$, and the other to IR CFT at $y=\infty$ with AdS length $\ell_{\text{AdS}}=1$. 
Hence conformal symmetry is broken in the boundary field theory for non-zero $\beta^+\beta^-$.  
\item When we set $\beta^+=-\beta^- \equiv \lambda $ and take the limit $\lambda \rightarrow \infty$, 
the term proportional to $y^{-4}dx^+dx^-$ dominates over $y^{-2} dx^+dx^-$ and the hypersurface $\Sigma_{\alpha\beta\gamma}$
asymptotes to a new AdS$_3$ vacuum with AdS length $\ell'_{\text{AdS}}=\frac{1}{2}\ell_{\text{AdS}}$. 
\end{itemize}

It was observed in \cite{NS} that the parameters $\beta^{\pm}$ play the role of flow  parameters of renormalization group. To identify this flow in the bulk  let us set $\beta^+=-\beta^- \equiv \beta$ in (\ref{per1}) for simplicity. 
\begin{equation}
ds_0^2\Big|_{\Sigma_{\alpha\beta\gamma}} = y^{-2}dy^2-y^{-4}(y^2 \cosh 2\gamma+4\beta^2) dx^+dx^- \label{fmetric}
\end{equation}
In general, in the bulk of an asymptotically AdS space-time AdS symmetry is broken at $y \neq 0$ and conformal symmetry is also broken on the holographic screen located at this value of $y$. As $y$ gets closer to 0, then AdS symmetry will be recovered and the field theory on the holographic screen will flow in the UV to a fixed point, if it exists. 
Now, to describe a new type of flow on the holographic screen, we should change the radial variable $y$ to a new one $z$ as $y=2\sqrt{\beta z}$. Then the metric (\ref{fmetric}) is transformed to 
\begin{equation}
ds_0^2\Big|_{\Sigma_{\alpha\beta\gamma}}  = \frac{1}{4z^2}dz^2-\Big(\frac{\cosh 2\gamma}{y^2}+\frac{1}{4z^2}\Big)dx^+dx^-.  \label{flowtoW32}
\end{equation}
If $y$ and $\gamma$ are fixed, this is an asymptotically AdS metric, where $z$ is a new radial coordinate. 
In this case we can  consider a flow on a constant-$y$ holographic screen by fixing the value of $y$ and sending $z \rightarrow 0$. Then the field theory  flows to a UV fixed point and the above metric flows to that of AdS$_3$ with AdS length $ \ell_{\text{AdS}}'$.  Along the way $\beta$ goes to $\infty$. This flow is depicted in Fig.1.

\begin{figure}[thb]
\hspace {.1cm}     \begin{minipage}{1.\hsize}
     \begin{center}
\includegraphics[scale=.4, clip]{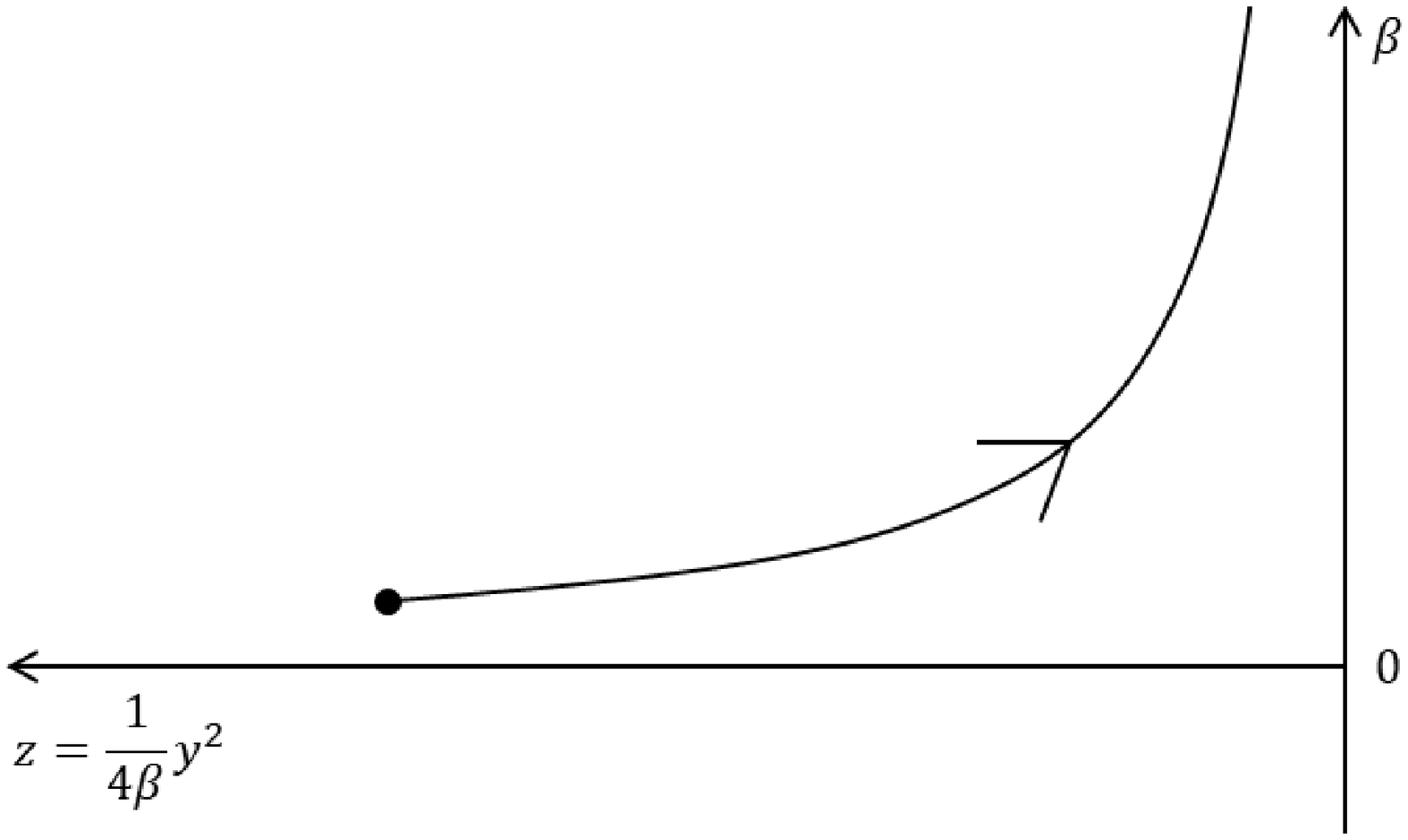}
\hspace{.2cm} 
        \end{center}
\hspace{1cm} Fig.1:   Flow in a $z$-$\beta$ plane along a curve $4\beta z=y^2$ with fixed $y$.
     \end{minipage}
\end{figure}

On the CFT side this flow is associated with a change of the translation operator $L^h_{-1} \rightarrow -(1/4)W^h_{-2}$ as follows.\cite{NS} 
On the boundary of  a chosen hypersurface $\Sigma_{\alpha\beta\gamma}$ (holomorphic) primary operators in general have a form,
\begin{equation}
e^{ix^+ L^h_{-1}} e^{i\alpha^+ W^h_{-2}} e^{\beta^+ W^h_{-1}}  {\cal O}(0) e^{-\beta^+ W^h_{-1}} e^{-i\alpha^+ W^h_{-2}}e^{-ix^+ L^h_{-1}}.
\end{equation} 
Let us concentrate on (global) Virasoro symmetry, because $\alpha$, $\beta$ and $\gamma$ are fixed. For simplicity we set $\gamma=0$ here. 
When  the correlation functions of scalar operators on the common boundary of $\Sigma_{\alpha\beta\gamma}$ are computed,  they depend on $\beta^+$ in addition to $x$, because the exponentials $\exp\{ ix L^h_{-1}\} $ and $ \exp\{\beta W^h_{-1}\}$ do not commute. They do not depend on $\alpha^+$, because $[L^h_{-1}, W^h_{-2}]=0$. For simplicity $\alpha^+$ will be omitted in the following discussion. Then an operator ${\cal O}(x^+,\beta^+)$ on the boundary are rewritten as 
\begin{eqnarray} 
{\cal O}(x^+,\beta^+) &\equiv& e^{ix^+ L^h_{-1}}e^{\beta^+ W^h_{-1}} {\cal O}(0) e^{-\beta^+ W^h_{-1}} e^{-ix^+ L^h_{-1}} \nonumber \\
&=&e^{\beta^+ W^h_{-1}} \bm{\cal O}(x)e^{-\beta^+W^h_{-1}},
\end{eqnarray}
where 
\begin{eqnarray}
\bm{\cal O}(x) &\equiv& e^{ix^+\bm{ L}^h_{-1}} {\cal O}(0) e^{-ix^+\bm{L}^h_{-1}}, \label{bmO} \\
\bm{L}^h_{-1} &\equiv& e^{-\beta^+W^h_{-1}}  L^h_{-1} e^{\beta^+W^h_{-1}}  =L^h_{-1}-\beta^+ W^h_{-2}. \label{bmL}
\end{eqnarray}
In the limit $\beta^+ \rightarrow \infty$ (\ref{bmL}) asymptotes to $-\beta^+ W^h_{-2} $, which is proportional to the Virasoro generator $\hat{L}_{-1}=-(1/4) W^h_{-2} $ in the ${\cal W}_3^{(2)}$ algebra\cite{Kraus}\cite{Castro}. And the central charge of ${\cal W}_3^{(2)}$ CFT is $c/4$. Hence up to a congruence transformation, all operators of form ${\cal O}(x^+,\beta^+)$ are mapped onto operators of form (\ref{bmO}), where the translation operator is modified. In \cite{NS} by using an explicit form of the two-point function it was also shown that as  $\beta^{+} \rightarrow \infty$  two-point functions of scalar operators in the boundary CFT flow from that in $W_3$ vacuum to that in $W_3^{(2)}$ vacuum and the conformal weight changes from $h$ to $h/2$. This flow is generated by a perturbation $\Delta S=\beta^+ W^h_{-1}$ to the action.  This flow does not contradict  the c-theorem\cite{Zamo}, because $[W_1,W_{-1}]=-2L_0$ and the `metric' $G_{ij}(g)$ is not positive-definite in the Euclidean CFT.

In this paper we will show that holography of 3d spin-3 gravity and the boundary ${\cal W}_3$ CFT is realized in the auxiliary  8d space. This will be done in the following steps. In the above discussion Klein-Gordon equation is associated with the quadratic Casimir operator. In sec. 2 of this paper the other equation for the scalar field, which stems from the cubic Casimir operator of $su(1,2)$, is studied. 
 It is shown that the coefficient functions of the third-order derivatives in this third-order differential equation coincide with the spin-3 gauge field $\varphi_{\mu\nu\lambda}$, when restricted to 3d hypersurface $\Sigma_{\alpha\beta\gamma}$. 
In sec. 3 a bulk to boundary propagator is obtained by solving the Klein-Gordon equation and it is shown that semi-classically evaluated path integral for a free scalar field in the background (\ref{8dmetric}) yields a generating functional for a two-point function of scalar operators on the boundary.  In sec. 4 an 8d action integral  for the spin-3 gravity coupled to a scalar field is proposed. In sec. 5 black hole solutions, both with and without spin-3 charge, are obtained. These steps are carried out by introducing 8d local frame and $sl(3,\mathbb{R}) \times sl(3,\mathbb{R})$ flat connections, $A=\omega+e$ and $\overline{A}=\omega-e$. These are 8 $\times$ 8 matrices. By construction, on a 3d  hypersurface $\Sigma_{\alpha\beta\gamma}$, these flat connections reduce to $3 \times 8$ matrices and they yield new black hole solutions to the 3d $SL(3,\mathbb{R}) \times SL(3,\mathbb{R})$  Chern-Simons theory. The flat connections on $\Sigma_{\alpha\beta\gamma}$ have properties $A_{x^-}=0$ and $\overline{A}_{x^+}=0$. The integrability condition for the partition function of the charged black hole is checked. The partition function coincides with that of the solution obtained in \cite{Kraus}, although the flat connections of the two solutions  satisfy distinct boundary conditions. This paper is summarized in Sec. 6. In appendix A  conventions for $sl(3,\mathbb{R})$ and $su(1,2)$ algebras in this paper are presented. In appendix B a representation of ${\cal W}_3$ generators in terms of differential operators is  presented. In appendices C and D the explicit forms of Klein-Gordon equation for a scalar field $\Phi$ and a spin-3 field $\phi_{\mu\nu\lambda}$ in the auxiliary space is presented.
Black hole solutions are obtained by adding extra terms $\psi$ and $\overline{\psi}$ to the flat connections for $su(1,2) \times su(1,2)$ symmetric space-time.  In appendix E equations for their first-order perturbations  $\psi^{(1)}$ and $\overline{\psi}^{(1)}$ are presented and the solutions to them are shown. In appendices F, G and H results for black hole solutions with and without spin-3 charge are presented. 

\section{Equation Related to the Cubic Casimir Operator}
\hspace{5mm}
As explained in sec. 1 the local state for a scalar field  in the bulk $|\Phi(x^+,\alpha^+,\beta^+,x^-,\alpha^-,\beta^-,y,\gamma) \rangle$  satisfies an eigenstate equation for the quadratic Casimir operator. This state also satisfies an equation corresponding to  the cubic  Casimir:
\begin{equation}
\Big( \tilde{C}_3(L^h,W^h) + \tilde{C}_3(\overline{L}^h,\overline{W}^h) \Big)|\Phi\rangle= -\frac{3i}{2}q \Big((\Delta-4)^2+q^2 \Big) | \Phi\rangle,              \label{C3}
\end{equation}
where $h=\Delta/2$  and $q$ are the conformal weight and the spin-3 charge of the boundary primary state $|O_{\Delta,q}\rangle$. $\tilde{C}_3$ is defined in (\ref{C3tilde}).  
By substituting (\ref{diffL}), (\ref{differentialop}) into (\ref{C3}) a differential equation for a scalar field $\Phi$ in the bulk is obtained. After some calculation, this can be succinctly written in the following simple form.
\begin{equation}
\phi^{\mu\nu\lambda} \, \nabla_{\mu} \, \nabla_{\nu} \, \nabla_{\lambda} \, \Phi=-\frac{3i}{2}q \Big((\Delta-4)^2+q^2 \Big)\Phi,  
\label{thirdorder}
\end{equation}
where $\nabla_{\mu}$ is a covariant derivative for the metric (\ref{8dmetric}) with the Christoffel symbol, and $\phi^{\mu\nu\lambda}$ is a completely symmetric tensor. The explicit form of this field is given in Appendix D. The imaginary unit $i$ on the right hand side of (\ref{thirdorder}) implies that $\Phi$ is a complex function. (See the solution (\ref{b2bndry}).)

On a 3d hypersurface $\Sigma_{\alpha\beta\gamma}$ with constant $\alpha^{\pm}$, $\beta^{\pm}$ and $\gamma$, which was introduced in sec 1,  (\ref{phiCubic}) simplifies to
\begin{equation}
\phi \Big|_{\Sigma_{\alpha\beta\gamma}}=\frac{3}{4y^4} \, \Big(\beta^-dx^-(dx^+)^2+\beta^+dx^+(dx^-)^2-y\sinh 2\gamma \, dydx^+dx^-\Big).  \label{phisigma}
\end{equation}
This field breaks Lorentz symmetry on the boundary of $\Sigma_{\alpha\beta\gamma}$, on which the values of $\beta^{\pm}$ are fixed. It is also found that 8d covariant derivative of $\phi_{\mu\nu\lambda}$ vanishes.
\begin{equation}
\nabla_{\rho} \, \phi_{\mu\nu\lambda} = 0 \label{nabphi}
\end{equation}
This fact gives a geometrical meaning to the $\phi_{\mu\nu\lambda} $ field. 

It will soon be shown that $\phi_{\mu\nu\lambda}$  actually coincides with a spin-3 gauge field on a hypersurface $\Sigma_{\alpha\beta\gamma}$. This suggests that both Casimir equations have geometrical meaning via the metric and spin-3 gauge field. 
For this purpose we introduce a vielbein field $e^a_{\mu}$. Here $a=1,2, \ldots , 8$ and $\mu=x^{\pm}, \alpha^{\pm}, \beta^{\pm}, y, \gamma$. This is a local-frame field in 8 dimensions and is an 8 $\times$ 8 matrix. Notice that this is different from the 3 $\times $ 8 rectangular vielbein introduced in \cite{Campoleoni}. It is required that\footnote{Throughout this paper the vielbein $e^a_{\mu}$ is assumed to be invertible.} 
\begin{equation}
g_{\mu\nu}= \frac{1}{2} \, \text{tr} (e)^2=h_{ab} \, e^a_{\mu} \, e^b_{\nu}.  \label{ghee}
\end{equation}

To impose more restrictions we consider $sl(3,\mathbb{R})$ connections\footnote{See appendix A for our conventions for $sl(3,\mathbb{R})$ algebra. The reason for using $sl(3,\mathbb{R})$ generators $t_a$, not $su(1,2)$ ones $\tilde{t}_a$ is that if $su(1,2)$ generators are used, then it turns out that $e^a_{\mu}$ $(a=4 \sim 8)$ becomes pure imaginary.
If $sl(3,\mathbb{R})$ generators are used instead, $e^a_{\mu}$ is real. Hence the symmetry algebra of the local frame is $sl(3,\mathbb{R}) \times sl(3,\mathbb{R})$, while that of the space-time is $su(1,2) \times su(1,2)$. The vielbein connects two analytically continued spaces.}
\begin{eqnarray}
A &=& \omega+e=t_a \, (\omega^a_{\mu}+e^a_{\mu}) \, dx^{\mu}, \label{26}\\
\overline{A} &=& \omega-e=t_a \, (\omega^a_{\mu}-e^a_{\mu}) \, dx^{\mu}, \label{27}
\end{eqnarray}
and require flatness conditions on them.
\begin{eqnarray}
F &=& dA+ A \wedge A=0, \label{F8d}\\
\overline{F} &=& d\overline{A}+\overline{A} \wedge \overline{A}=0.  \label{Fbar8d}
\end{eqnarray}
Note that these are equations in 8d, and there are no Chern-Simons actions which classically lead to (\ref{F8d}) and  (\ref{Fbar8d}).\footnote{However, as will be discussed later,  there exists an 8d Einstein-like action such that its equations of motion coincide with (\ref{F8d}) and  (\ref{Fbar8d}), provided the metricity condition and invertibility of the vielbein are  assumed for the local frame fields.}  
On a hypersurface $\Sigma_{\alpha\beta\gamma}$ with constant $\alpha^{\pm}$, $\beta^{\pm}$ and $\gamma$, however, these frame fields reduce to 3 $\times$ 8 rectangular matrices, and 
these equations are nothing but the equations of motion for connections in $SL(3,\mathbb{R}) \times SL(3,\mathbb{R})$ Chern-Simons gauge theory.

It turns out there are two distinct solutions to (\ref{F8d}) and (\ref{Fbar8d}), which produce  (\ref{8dmetric}).
\begin{enumerate}
\item [$\bullet$ Solution (I)] 
\begin{eqnarray}
A &=& \frac{1}{y}dy \, t_2+\frac{1}{y}(\cosh\gamma \, dx^+-\sinh \gamma \, d\beta^+)\, t_3+\frac{1}{2}d\gamma \, t_6 \nonumber \\
&& +\frac{1}{y}(\sinh \gamma dx^+-\cosh \gamma \, d\beta^+) \, t_7 
 -\frac{1}{y^2} \, (\beta^-dx^--d\alpha^-) \, t_8, \label{Kraus1}\\
\overline{A} &=& -\frac{1}{y}dy \, t_2-\frac{1}{y}(\cosh \gamma \, dx^--\sinh \gamma d\beta^-) \, t_1-\frac{1}{2} \, d\gamma \, t_6 \nonumber \\
&& +\frac{1}{y^2} \, (\beta^+ \, dx^+-d\alpha^+) \, t_4-\frac{1}{y} \, (\sinh\, \gamma \, dx^--\cosh \gamma \, d\beta^-) \, t_5 \label{Kraus2}
\end{eqnarray}
We note that both $A_{x^{+}}$ and $A_{x^-}$ are non-vanishing. Similarly $\overline{A}_{x^{\pm}} \neq 0$. 
As $\beta^{\pm}$ increase from $0$ to $\infty$ in this solution, the leading terms interchange between $A_{x^+}$ and $A_{x^-}$. When this solution is restricted to a hypersurface $\Sigma_{\alpha\beta\gamma}$ (especially for $\gamma =0$), it  coincides with the interpolating solution eq (2.27) of \cite{Kraus}.

\item [$\diamond$ Solution (II)] 
\begin{eqnarray}
A &=& \frac{1}{y}dy \, t_2+\frac{1}{y}(\cosh\gamma \, dx^+-\sinh \gamma \, d\beta^+)\, t_3-\frac{1}{2}d\gamma \, t_6 \nonumber \\
&& +\frac{1}{y}(\sinh \gamma dx^+-\cosh \gamma \, d\beta^+) \, t_7 
 +\frac{1}{y^2} \, (\beta^+dx^+-d\alpha^+) \, t_8, \label{background1}\\
\overline{A} &=& -\frac{1}{y}dy \, t_2+\frac{1}{y}(-\cosh \gamma \, dx^-+\sinh \gamma d\beta^-) \, t_1+\frac{1}{2} \, d\gamma \, t_6 \nonumber \\
&& -\frac{1}{y^2} \, (\beta^- \, dx^--d\alpha^-) \, t_4+\frac{1}{y} \, (-\sinh\, \gamma \, dx^-+\cosh \gamma \, 
d\beta^-) \, t_5 \nonumber \\
&&   \label{background2}
\end{eqnarray}
In this solution $A$ has no `-- components', $A_{x^-}=A_{\alpha^-}=A_{\beta^-}=0$, while $\overline{A}$ has no `+ components'.   Then on a hypersurface $\Sigma_{\alpha\beta\gamma}$, we have $A_{x^-}=0$. This is a flat connection in $SL(3,\mathbb{R}) \times SL(3,\mathbb{R})$ Chern-Simons gauge theory with a boundary condition $A_{x^-}=0$.  
\end{enumerate}

For solution (II), it can be checked that (\ref{ghee}) with $e=(1/2)(A-\overline{A})$ coincides with the metric (\ref{8dmetric}). Furthermore spin-3 gauge field
\begin{equation}
\varphi = \frac{1}{3!} \,
 \text{tr} (e)^3  \label{spin3gaugefield}
\end{equation}
exactly coincides with $\phi$ in (\ref{phiCubic}) up to a multiplicative constant. On $\Sigma_{\alpha \beta \gamma}$ this coincides with (\ref{phisigma}). 

On the other hand in the case of solution (I),  although the metric agrees with (\ref{8dmetric}), the spin-3 gauge field (\ref{spin3gaugefield})
 does not coincide with (\ref{phiCubic}). For simplicity, only the result for $\varphi$ on a hypersurface $\Sigma_{\alpha \beta \gamma}$ is presented here.
\begin{eqnarray}
\varphi^{\text{solution I}}\Big|_{\Sigma_{\alpha\beta\gamma}} &=& -\frac{3}{2y^4}\{\beta^-(dx^-)^3+\beta^+(dx^+)^3\}+\frac{3}{2y^3} \, \sinh \gamma \, dydx^+dx^- 
\neq  \phi|_{\Sigma_{\alpha\beta\gamma}} \nonumber \\ &&\label{fake}
\end{eqnarray} 
On $\Sigma_0$ two solutions (I) and (II) coincide. Moving away from $\Sigma_0$, they do not agree.

To conclude, out of the two sets of flat connections  (I) and (II), both of which are invariant under $SU(1,2) \times SU(1,2)$ and produce the same metric (\ref{8dmetric}), only for flat connections (II) the coefficient function $\phi_{\mu\nu\lambda}$ (\ref{phiCubic}) in the cubic-order differential equation (\ref{thirdorder}) for a scalar field coincides with the spin-3 gauge field (\ref{spin3gaugefield}). This is in accord with the fact that the coefficient function $g^{\mu\nu}$ of derivatives in the Klein-Gordon equation is the metric field. The two equations for the scalar field are written in terms of the geometrical quantities.  On the boundary of $\Sigma_{\alpha \beta \gamma}$ this solution satisfies  boundary conditions $A_{x^-}=\overline{A}_{x^+}=0$. The connection (II) on $\Sigma_{\alpha \beta \gamma}$ is the interpolating solution between IR and UV. 
 In sec. 5 we will construct black hole solutions by extending this solution.

\section{Bulk-to-Boundary Propagator}
\hspace{5mm}
In this section a bulk-to-boundary propagator for a scalar field $\Phi$ propagating in the space-time (\ref{8dmetric}) is derived and then a generating functional of a two-point function of scalar operators on the boundary is obtained semi-classically 
by substituting the solution to the equation of motion into the action of a scalar field.

Construction of the bulk-to-boundary propagator  is carried out by solving Klein-Gordon equation (\ref{KG}). First of all, in the boundary limit it should asymptote to the two-point function of scalar operators on the boundary, which was derived in \cite{NS}\cite{FateevRibault}.
\begin{eqnarray}
&&\langle O_{\Delta, q}(x_1^+,\alpha_1^+, \beta_1^+, x_1^-,\alpha_1^-,\beta_1^-) \, O_{\Delta,q}^{\dagger}(x_2^+,\alpha_2^+, \beta_2^+, x_2^-,\alpha_2^-,\beta_2^-)\rangle \nonumber \\
&=& \Big( D_{12} \, \overline{D}_{12}\Big)^{-(\Delta+3iq)/4} \, \Big( D_{12}^{\ast} \, \overline{D}_{12}^{\ast}\Big)^{-(\Delta-3iq)/4}, \label{twopoint}
\end{eqnarray}
where $D_{12}$, {\em etc} are defined by
\begin{eqnarray}
D_{12} &=& (x_{12}^+)^2-(\beta_{12}^+)^2-2x_{12}^+(\beta_1^++\beta_2^+)+4\alpha_{12}^+, \nonumber \\
\overline{D}_{12} &=& (x_{12}^-)^2-(\beta_{12}^-)^2-2x_{12}^-(\beta_1^-+\beta_2^-)+4\alpha_{12}^-, \nonumber \\
D_{12}^{\ast} &=& (x_{12}^+)^2-(\beta_{12}^+)^2+2x_{12}^+(\beta_1^++\beta_2^+)-4\alpha_{12}^+, \nonumber \\
\overline{D}^{\ast}_{12} &=& (x_{12}^-)^2-(\beta_{12}^-)^2+2x_{12}^-(\beta_1^-+\beta_2^-)-4\alpha_{12}^-. \label{DDstar}
\end{eqnarray}
Here $x_{12}^+ \equiv x_1^+-x_2^+$ {\em etc}.

Solution $\Phi=K_{\Delta,q} $ is obtained by power series expansion in $y$ near the boundary $y \sim 0$\footnote{This expansion is carried out in a region, where $D_{12}, \cdots \neq 0$. }: 
\begin{multline}
K_{\Delta,q}  (y, x_1^+,\alpha_1^+, \beta_1^+, x_1^-,\alpha_1^-,\beta_1^-,\gamma_1; x^+_2,\alpha_2^+, \beta_2^+, x_2^-,\alpha_2^-,\beta_2^- ) \\
= y^{\Delta} \, f_0(x_1^+,\alpha_1^+, \beta_1^+, x_1^-,\alpha_1^-,\beta_1^-) \, e^{-iq \, \gamma_1} +y^{\Delta+2} \, f_1(x_1^+,\alpha_1^+, \beta_1^+, x_1^-,\alpha_1^-,\beta_1^-,\gamma_1)  \\
+y^{\Delta+4} \, f_2(x_1^+,\alpha_1^+, \beta_1^+, x_1^-,\alpha_1^-,\beta_1^-,\gamma_1) + O(y^{\Delta+6}) \label{solKG}
\end{multline}
Here $( y, x_1^+,\alpha_1^+, \beta_1^+, x_1^-,\alpha_1^-,\beta_1^-,\gamma_1 )$ is a bulk point and $( x_2^+,\alpha_2^+, \beta_2^+, x_2^-,\alpha_2^-,\beta_2^- )$ a boundary point.
The first term in (\ref{solKG}) should  solve (\ref{KG}) to the leading order and the function $f_0$ must be chosen to be
\begin{equation}
f_0 = \Big( D_{12} \, \overline{D}_{12}\Big)^{-(\Delta+3iq)/4} \, \Big( D_{12}^{\ast} \, \overline{D}_{12}^{\ast}\Big)^{-(\Delta-3iq)/4},
\end{equation}
due to the boundary condition. The exponential factor  for the bulk point is introduced to the first term in (\ref{solKG}). By substituting this solution into (\ref{KG}) an equation for $f_1$ is obtained and it is readily solved. By repeating this procedure we get the following solutions 
\begin{eqnarray}
f_1 &=& \frac{1}{2} \Big\{ (\Delta+3iq)\frac{(x^+_{12}+\beta^+_{12})(x_{12}^-+\beta_{12}^-)}{D_{12} \, \overline{D}_{12}} \, e^{(-iq-2) \, \gamma_1} \nonumber \\
&&+(\Delta-3iq)\frac{(x^+_{12}-\beta^+_{12})(x_{12}^--\beta_{12}^-)}{D_{12}^{\ast} \, \overline{D}^{\ast}_{12}} \, e^{(-iq+2) \, \gamma_1} \Big\} \, f_0,
\end{eqnarray}
\begin{eqnarray}
f_2 &=& \Big\{ \frac{(\Delta+3iq)(\Delta+3iq+4)}{8} \, \frac{(x^+_{12}+\beta^+_{12})^2 \, (x_{12}^-+\beta^-_{12})^2}{(D_{12}\overline{D}_{12})^2}
e^{-4\gamma_1} \nonumber \\
&& + \frac{(\Delta-3iq)(\Delta-3iq+4)}{8} \, \frac{(x^+_{12}-\beta^+_{12})^2 \, (x_{12}^--\beta^-_{12})^2}{(D^{\ast}_{12}\overline{D}^{\ast}_{12})^2}e^{4\gamma_1} \nonumber \\
&& +\frac{\Delta^2-4\Delta+12iq+9q^2}{16} \, \frac{1}{D^{\ast}_{12}\overline{D}^{\ast}_{12}}+\frac{\Delta^2-4\Delta-12iq+9q^2}{16} \, \frac{1}{D_{12}\overline{D}_{12}} \nonumber \\
&&+ \frac{\Delta^2+9q^2}{16} \, \Big(\frac{1}{\overline{D}_{12} \, D^{\ast}_{12}}+\frac{1}{D_{12} \, \overline{D}^{\ast}_{12} } \Big) \, \Big\}f_0 e^{-iq \gamma_1}.
\end{eqnarray}
Now up to order $y^{\Delta+4}$ the series (\ref{solKG}) can be  summed up with the following result. 
\begin{eqnarray}
K_{\Delta,q} &=& y^{\Delta} \, e^{-iq \gamma_1} \, \Big\{ D_{12}\overline{D}_{12}-2y^2(x^+_{12}+\beta^+_{12})(x^-_{12}+\beta^-_{12})e^{-2\gamma_1}+y^4\Big\}^{-(\Delta+3iq)/4} \nonumber \\
&& \Big\{ D^{\ast}_{12}\overline{D}^{\ast}_{12}-2y^2(x^+_{12}-\beta^+_{12})(x^-_{12}-\beta^-_{12})e^{2\gamma_1}+y^4\Big\}^{-(\Delta-3iq)/4} \label{b2bndry}
\end{eqnarray}
It is directly checked that (\ref{b2bndry}) solves Klein-Gordon equation (\ref{KG}) exactly.
Furthermore, it is checked that this propagator also satisfies up to order $y^{\Delta+4}$ the equation (\ref{thirdorder}) which is related to the cubic Casimir operator.     As an independent check, we also found that conditions for $su(1,2) \times su(1,2)$ invariance of the propagator are satisfied: $\lim_{\, y_2 \rightarrow 0} \, (L^{(1)}_n+L^{(2)}_n) \, \Big(y_2^{\Delta} \, e^{iq\gamma_2}\, K_{\Delta,q}\Big)=0$ and similar equations containing $W^{(i)}_n$ are satisfied, where $L^{(i)}_n$ and $W^{(i)}_n$ are differential operators defined in (\ref{diffL}), (\ref{differentialop}), which act on the $i$-th variable. Hence it is established  that this bulk-to-boundary propagator is an exact solution.

By using the bulk-to-boundary propagator a scalar field inside the bulk is reconstructed in terms of a boundary CFT operator.
This provides a more explicit expression for the extrapolating dictionary than that of the local state for a scalar field obtained in eq (3.9) of \cite{NS}. 
\begin{multline}
\Phi(y,x^+,\alpha^+,\beta^+,x^-,\alpha^-,\beta^-,\gamma) \\
= \int d^2 x_2 d^2\alpha_2 d^2 \beta_2 \, K_{\Delta,q}(y,x^+,\alpha^+, \beta^+, x^-,\alpha^-,\beta^-,\gamma; x^+_2,\alpha_2^+, \beta_2^+, x_2^-,\alpha_2^-,\beta_2^-) \\
 O_{\Delta,q}(x^+_2,\alpha_2^+, \beta_2^+, x_2^-,\alpha_2^-,\beta_2^-)
\end{multline}
All $W$-descendants of the primary scalar operators correspond to the scalar field in the bulk. 

Now let us switch to a Euclidean space by a Wick rotation.
\begin{eqnarray}
x^+ &=& z, \quad x^-=-\bar{z}, \quad 
\beta^+= i\xi, \quad \beta^-= i\bar{\xi}, \quad 
\alpha^+=i\zeta, \quad \alpha^-=-i\bar{\zeta}  \label{Wick}
\end{eqnarray}
Here $\bar{z}$, $\bar{\xi}$ and $\bar{\zeta}$ are complex conjugates of $z$, $\xi$, $\zeta$.
The metric (\ref{8dmetric}) becomes
\begin{eqnarray}
ds_0^2 &=& \frac{1}{y^{2}} \, dy^2+\frac{1}{3}d\gamma^2+\frac{4}{y^4} \, |d\zeta-\xi \, dz|^2 \nonumber \\
&&+\frac{1}{y^2} \cosh(2\gamma) \big|dz-i \tanh(2\gamma) \, d\xi \big|^2+\frac{1}{y^2}\frac{1}{\cosh (2\gamma)} \, |d\xi|^2
\end{eqnarray}
It can be shown that in the  region including $z_{12}, \, \xi_{12}, \, \zeta_{12}  \sim 0$ and in the $y \rightarrow 0$
limit $K_{\Delta,q}$ behaves as 
\begin{eqnarray}
K_{\Delta,\mu} & =& {\cal N}(\gamma_1) \, y^{8-\Delta} \, e^{-iq \gamma_1} \delta^2(z_{12}) \, \delta^2(\xi_{12}) \, \delta^2(\zeta_{12})+\cdots \nonumber \\
&&+ y^{\Delta} \, e^{-iq \gamma_1} \,  (D^E_{12}\overline{D}^E_{12})^{-(\Delta+3iq)/4} \, (D^{E\ast}_{12} \overline{D}^{E\ast}_{12})^{-(\Delta-3iq)/4}+\cdots \label{expK}
\end{eqnarray}
Here  the dots stand for terms with higher order powers of $y$, and $D^E_{12}, \ldots$ are obtained by replacing variables in $D_{12}, \ldots$ according to the rule of analytic continuation (\ref{Wick}). ${\cal N}(\gamma)$ is a function of $\gamma$:
\begin{eqnarray}
{\cal N}(\gamma) &=& \int d^2z d^2\xi d^2\zeta \, (1+2e^{-2\gamma} |z+i\xi|^2+|z^2+\xi^2-2iz\xi+4i\zeta|^2)^{-(\Delta+3iq)/4} \nonumber \\
&& ( 1+2e^{2\gamma} |z-i\xi|^2+|z^2+\xi^2+2iz\xi-4i\zeta|^2)^{-(\Delta-3iq)/4}.
\end{eqnarray}

We define action integral for a scalar field coupled to spin-3 gravity  by 
\begin{eqnarray}
S_{\text{scalar}} &=& \int dy \, d^2z d^2\xi d^2\zeta d\gamma \, \sqrt{g} \, \Big( 
\nabla \Phi^{\ast} \, \nabla \Phi +m^2 \, \Phi^{\ast} \, \Phi \Big). \label{scalaraction}
\end{eqnarray}
When a solution to the Klein-Gordon equation
\begin{eqnarray}
\Phi (y, z,\xi,\zeta,\gamma)
&=& \int d^2z' d^2\xi' d^2\zeta' \, K_{\Delta,q}(y,z,\xi,\zeta,\gamma; z',\xi',\zeta') \, \phi(z',\xi',\zeta'),
\end{eqnarray}
where $\phi(z,\xi,\zeta)$ is a boundary condition,  
is substituted into (\ref{scalaraction}) and $\sqrt{g}=y^{-9}/(2\sqrt{3})$ is used, a generating functional for the two-point function is obtained as a surface integral on the boundary by using the standard method\cite{Witten}. \footnote{Note that there is no delta function for $\gamma$ in the expansion (\ref{expK}) and the $\gamma$ integral factors out in (\ref{generating}).}
\begin{eqnarray}
S_{\text{scalar}} [\phi,\phi^{\ast}]&\propto& \Big(\int^{\infty}_{-\infty} d\gamma \, {\cal N}(\gamma)e^{-iq \gamma} \Big) \, \int d^2z_1 d^2 \xi_1 d^2 \zeta_1 \int d^2z_2 d^2 \xi_2 d^2 \zeta_2\nonumber \\
&&  \ \ 
\phi^{\ast} (z_1,\xi_1,\zeta_1,\bar{z}_1,\bar{\xi}_1,\bar{\zeta}_1) 
\quad (D^E_{12}\overline{D}^E_{12})^{-(\Delta+3iq)/4} \nonumber \\ && \qquad (D^{E\ast}_{12}\overline{D}^{E\ast}_{12})^{-(\Delta-3iq)/4} \, 
\phi(z_2,\xi_2,\zeta_2 \bar{z}_2,\bar{\xi}_2,\bar{\zeta}_2) \label{generating}
\end{eqnarray}
Hence both holographic dictionaries of AdS/CFT, the WGKP\cite{Witten}\cite{GKP} and BDHM dictionaries\cite{BDHM},  also hold in the case of 3d spin-3 gravity which couples to a scalar field.

\section{Action Integral for  3D Spin-3 Gravity Coupled to a Scalar Field}
\hspace{5mm}
One of the purpose of  this paper is to find out a formulation of 3d spin-3 gravity coupled to a scalar field in terms of 8d auxiliary  bulk space-time. A natural action integral for the scalar field (\ref{scalaraction}) was found in sec. 3. To make this formulation complete, it is necessary to write down an action for the gravity sector in the 8d auxiliary  space-time. 

In (\ref{26}) and (\ref{27})  we introduced 8d vielbein  $e^a_{\mu}$, spin connection $\omega^a_{\mu}$ and gauge connections $A=\omega +e$ and $\overline{A}=\omega-e$. By solving flatness conditions for the connections (\ref{F8d}) and (\ref{Fbar8d}) solutions to  these connections are obtained. If the space-time is restricted to a hypersurface with constant $\alpha^{\pm}$, $\beta^{\pm}$ and $\gamma$, the connections also become solutions to equations of motion of 3d $SL(3,\mathbb{R}) \times SL(3,\mathbb{R})$ Chern-Simons gauge theory. 

In 8d space-time Chern-Simons gauge theory does not exist, and it is not possible to formulate the gravity sector in a similar way. It is, however, possible to write down an 8d action which reproduces the same equations of motion for the vielbein at the semi-classical level.

Let us write the flatness condition for the connections (\ref{F8d}), (\ref{Fbar8d}) in terms of $\omega^a$ and $e^a$:
\begin{eqnarray}
&& de^a+{f^a}_{bc} \, \omega^b \wedge e^c =0, \label{de} \\
&& d\omega ^a+\frac{1}{2} \, {f^a}_{bc} \, \omega^b \wedge \omega^c +\frac{1}{2} \, {f^a}_{bc} \, e ^b \wedge e^c=0. \label{domega}
\end{eqnarray}
These are linear combinations of the flatness conditions. 
In terms of components the first equation is given by a torsionless condition
\begin{equation}
\nabla_{\mu} \, e^a_{\nu}+{f^a}_{bc} \, \omega^b_{\mu} \, e^c_{\nu}= \nabla_{\nu} \, e^a_{\mu}+{f^a}_{bc} \, \omega^b_{\nu} \, e^c_{\mu}.  \label{T=0}
\end{equation}
Here $\nabla_{\mu}$ is a covariant derivative for $g_{\mu\nu}=e^a_{\mu} \, e_{a\nu}$. 
If we restrict solutions to (\ref{T=0}) to satisfy the vielbein postulate, 
\begin{equation}
(D_{\mu}\, e_{\nu})^a=\nabla_{\mu} \, e^a_{\nu}+{f^a}_{bc}\omega_{\mu}^be^c_{\nu}=0, \label{S=0}
\end{equation}
which states that the full covariant derivative of $e^a_{\mu}$ should vanish,\footnote{This is equivalent to the condition of metricity, $\nabla_{\mu} \, g_{\nu\lambda}=0$. Furthermore all the spin-3 gauge field $\varphi_{\mu\nu\lambda}$ obtained in this paper, (\ref{phiCubic}) and those for black hole solutions, satisfy $\nabla_{\mu}\varphi_{\nu\lambda\rho}=0$.  This implies that (\ref{S=0}) is obeyed, because $\varphi_{\mu\nu\lambda}$ is proportional to $\text{tr}(e_{\mu}\{e_{\nu}, e_{\lambda}\})$.}
then by using (\ref{hff}) the spin connection $\omega^a_{\mu}$ is expressed in terms of $e^a_{\mu}$,
\begin{equation}
\omega^a_{\mu} = \frac{1}{12}{f^a}_{bc}\,  e^b_{\nu}\, \nabla_{\mu}e^{c\nu} \equiv \tilde{\omega}^a_{\mu}(e).
\end{equation}

We define the field strength for the gauge field of local frame $sl(3,\mathbb{R})$ transformation.
\begin{equation}
{R^a}_{\mu\nu}(\omega^a_{\lambda}) =
\partial_{\mu} \omega^a_{\nu}-\partial_{\nu}\omega^a_{\mu}+{f^a}_{bc} \, \omega^b_{\mu} \, \omega^c_{\nu}
\end{equation}
The second equation of the flatness condition (\ref{domega}) is now written as 
\begin{equation}
{R^a}_{\mu\nu}(\tilde{\omega}(e)) +{f^a}_{bc}e^b_{\mu} \, e^c_{\nu}=0. 
\end{equation}
This equation can be derived from the following action.
\begin{equation}
S_{\text{spin-3 gravity}} = \int d^8x \, |e| \,  \frac{1}{16\pi G} \, \{{f_a}^{bc} \, e^{\mu}_b \, e^{\nu}_c \, {R^a}_{\mu\nu}(\tilde{\omega}(e))+\frac{10}{3}\Lambda_8 \}, 
\label{s3g}
\end{equation}
where $|e|=\text{det} (e^a_{\mu})=\frac{3}{32}\sqrt{- g}$ is a determinant of the vielbein and $\Lambda_8$ is the cosmological constant in (\ref{cc}). $G$ is a Newton constant. Equivalence of the equations of motion can be shown by using the Jacobi's identity for $f_{abc}$. 
Because this is written in terms of the vielbein fields, it would be better, if it could  be rewritten in terms of fields without local-frame indices, such as $g_{\mu\nu}$ and $\varphi_{\mu\nu\lambda}$. 
Under $SL(3,\mathbb{R})$ gauge transformation $g_{\mu\nu}$ and $\varphi_{\mu\nu\lambda}$ transform as tensors of 8d diffeomorphism. 
In the similar case  of $SL(2,\mathbb{R}) \times SL(2,\mathbb{R})$ Chern-Simons theory the corresponding action coincides with an Einstein-Hilbert action with a cosmological constant.\cite{AT}\cite{W} Rewriting of  the action (\ref{s3g})  in terms of $g_{\mu\nu}$ and  $\varphi_{\mu\nu\lambda}$ may be achieved  through a perturbation theory around the background (\ref{background1}),  (\ref{background2}).  This will not be attempted in this paper. 
On the hypersurface with constant $\alpha^{\pm}$, $\beta^{\pm}$, $\gamma$, solutions to the equations of motion for (\ref{s3g}) reduce to  those in 3d spin-3 gravity represented by $SL(3,\mathbb{R}) \times SL(3,\mathbb{R})$ Chern-Simons theory. 

Action for a free charged scalar field is given by 
\begin{equation}
S_{\text{scalar}} = \int d^8x \, \sqrt{-g} \, \{ -\nabla^{\mu}\Phi^{\ast} \, \nabla_{\mu}\Phi-m^2 \, \Phi^{\ast}\Phi\}.
\end{equation}
The equation of motion is Klein-Gordon equation. 
In sec. 2 it was shown that the solution to  Klein-Gordon equation in the metric (\ref{8dmetric}) also satisfies the cubic-order differential equation.  Hence the Klein-Gordon equation is sufficient to determine the classical solution at least for this case. Although it is not clear if a cubic-order differential equation exists in other space-times such as black hole space-times, where there is no symmetry algebra to yield a set of Casimir operators, the quadratic-order differential equation will be also sufficient to determine the solution.  The complex scalar field $\Phi$ is assumed to have spin-3 charge $q$. Then the scaling dimension $\Delta$ of $\Phi$ is determined by (\ref{massDelta}). 
The self coupling of scalar fields can be introduced straightforwardly.  We propose that the total action
\begin{equation}
S_{\text{total}} = S_{\text{spin-3 gravity}}+S_{\text{scalar}}
\end{equation}
describes the 3d spin-3 gravity theory coupled to a scalar field. 
In the limit of large central charge ($G \rightarrow 0$), the action for spin-3 gravity determines the geometry of the 8d space semi-classically, and $S_{\text{scalar}}$ describes the scalar field in this background.

\section{Black Hole Solutions}
\hspace{5mm}
In this section perturbations around the background solution (\ref{background1}) and (\ref{background2}) are considered and new solutions to the flatness conditions, black hole solutions,  are obtained. 
First we will consider flat connections $A$ (\ref{background1}) and $\overline{A}$ (\ref{background2}) at $y=1$, which will be denoted as ${\cal A}_0$ and ${\cal \overline{A}}_0$, respectively.
\begin{eqnarray}
{\cal A}_0 &=& (\cosh\gamma \, dx^+-\sinh \gamma \, d\beta^+)\, t_3-\frac{1}{2}d\gamma \, t_6 \nonumber \\
&& +(\sinh \gamma dx^+-\cosh \gamma \, d\beta^+) \, t_7 
 + (\beta^+dx^+-d\alpha^+) \, t_8, \label{gau1}\\
{\cal \overline{A}}_0 &=&  (-\cosh \gamma \, dx^-+\sinh \gamma d\beta^-) \, t_1+\frac{1}{2} \, d\gamma \, t_6  \nonumber \\
&& - (\beta^- \, dx^--d\alpha^-) \, t_4+ (-\sinh\, \gamma \, dx^-+\cosh \gamma \, 
d\beta^-) \, t_5 \label{gau2}
\end{eqnarray}
Gauge connections at an arbitrary value of $y$ are obtained by carrying out the following gauge transformations
\begin{eqnarray}
A_0 &=& b(y)^{-1} {\cal A}_0 \, b(y)+b(y)^{-1} \, db(y), \label{gauge1} \\
\overline{A}_0 &=& b(y) {\cal \overline{A}}_0 b(y)^{-1}+b(y) \, db(y)^{-1},  \label{gauge2}
\end{eqnarray}
where $b(y) = y^{t_2}$. ${\cal A}_0$ and ${\cal \overline{A}}_0$ also satisfy the flatness conditions.
\begin{eqnarray}
{\cal F}_0 &=&d  {\cal A}_0+ {\cal A}_0 \wedge {\cal A}_0=0, \label{flat1}\\
{\cal \overline{F}}_0 &=& d{\cal \overline{A}}_0+{\cal \overline{A}}_0 \wedge {\cal \overline{A}}_0=0 \label{flat2}
\end{eqnarray}

Now we add small perturbations $\psi$ and $\overline{\psi}$ to ${\cal A}_0$ and ${\cal \overline{A}}_0$, respectively:
\begin{eqnarray}
{\cal A}&=& {\cal A}_0+\psi, \label{AP1}\\
{\cal \overline{A}} &=& {\cal \overline{A}}_0+\overline{\psi}.\label{AP2}
\end{eqnarray}
Then we impose conditions that ${\cal A}$ and ${\cal \overline{A}}$ should satisfy the flatness conditions (\ref{flat1}), (\ref{flat2}). Finally, gauge transformations $b(y)$ are used to obtain connections $A$ and $\overline{A}$. 

When $\psi$ and $\overline{\psi}$ are expanded as $\psi=\psi^{(1)}+\psi^{(2)}+\cdots$ and similarly for $\overline{\psi}$, 
where $\psi^{(i)}$ is an infinitesimal one-form at $i$-th order of perturbation, the flatness conditions to first order read
\begin{eqnarray}
&&d\psi^{(1)}+{\cal A}_0 \wedge \psi^{(1)}+\psi^{(1)} \wedge {\cal A}_0=0, \label{p1} \\
&& d\overline{\psi}^{(1)}+{\cal \overline{A}}_0 \wedge \overline{\psi}^{(1)}+\overline{\psi}^{(1)} \wedge {\cal \overline{A}}_0  =0. \label{p2}
\end{eqnarray}
These conditions will be solved explicitly.   By expanding $\psi^{(1)}$ and ${\cal A}_0$ into a basis of $sl(3,\mathbb{R})$ generators as
\begin{eqnarray}
\psi^{(1)}&=& \psi^{(1)a} \, t_a \\
{\cal A}_0 &=& {\cal A}_0^a \, t_a,
\end{eqnarray}
where summation over $a=1, \cdots, 8$ is not shown explicitly, eq (\ref{p1}) is transformed into
\begin{equation}
d\psi^{(1)a}+{f^a}_{bc}\, {\cal A}_0^b \wedge \psi^{(1)c}=0. \label{flat+} 
\end{equation}
Explicit forms of these equations are presented in (\ref{E1}) in appendix E. 
For example, for $a=4$ the equation is simple:
$d\psi^{(1)4}=0$. 
This is trivially solved locally: $\psi^{(1)4}=-dQ_1$, where $Q_1$ is an arbitrary function. Other equations are also solved by taking appropriate linear combinations of (\ref{flat+}) which give closed  forms. General  solutions are presented in (\ref{psia}) and those for $\overline{\psi}^{(i)a}$ in (\ref{psibara}) in appendix E.
These solutions contain 16 arbitrary $Q_n$, $\overline{Q}_n$. Actually, these are gauge modes. However, when  those modes are changed by amounts which are not single-valued on the torus, then the flat connections before and after the change are inequivalent. For static or stationary black hole solutions the functions $Q_n$ and $\overline{Q}_n$ have to be chosen such that $\psi^{(1)a}$ and $\overline{\psi}^{(1)a}$ are periodic in variables  $x^{\pm}$.

\subsection{Asymptotically AdS$_3$  Black Hole Solutions without Spin-3 Charge}
\hspace{5mm}

The flat connections for a black hole without spin-3 charge are obtained by choosing suitable $Q_n$ and $\overline{Q}_n$, which yield static or stationary connections. The exact terms $dQ_n$, $d\overline{Q}_n$ in (\ref{psia}), (\ref{psibara}) are determined to make $\psi^{(1)a}$ and $\overline{\psi}^{(1)a}$ periodic  in  $x^{\pm}$. 
The results for $Q_n$ are presented in (\ref{Qno-1}).
The results for $\psi^{(1)a}$ and $\overline{\psi}^{(1)a}$ are also presented in (\ref{psino1}) in appendix F.
In these results parameters $a$ and $\bar{a}$ are the following constants. 
\begin{eqnarray} 
a &=&  2G \, (M+J), \label{a}\\
\bar{a}  &=& 2G \, (M-J) \label{bara}
\end{eqnarray}
Here $M$ and $J$ are mass and angular momentum, $G$ a Newton constant.   

Now, although these results are first-order perturbations in the parameters $a$ and $\bar{a}$,  (\ref{AP1}), (\ref{AP2}) with $\psi=\psi^{(1)}$ and $\overline{\psi}=\overline{\psi}^{(1)}$ satisfy the full flatness conditions. This is because $\psi^{(1)}$ and $\overline{\psi}^{(1)}$ are proportional to $dx^+$ and $dx^-$, respectively, and $\psi^{(1)} \wedge \psi^{(1)}=\overline{\psi}^{(1)} \wedge \overline{\psi}^{(1)}=0$ hold. 
Hence these are exact solutions. The gauge connections $A$, $\overline{A}$ with $y$ components are obtained by the gauge transformations (\ref{gauge1}) and (\ref{gauge2}). The vielbein $e=\frac{1}{2}\, (A-\overline{A})$ then yields the metric $g_{\mu\nu}= (1/2) \text{tr} (e)^2$. The metric tensor for the 8d space-time is presented in (\ref{btz}). 
It turns out this metric satisfies 8d vacuum Einstein equation
\begin{equation}
R_{\mu\nu}-\frac{1}{2} g_{\mu\nu}R=-\Lambda_8 \, g_{\mu\nu} \label{cc}
\end{equation}
with $\Lambda_8=-36$. 

 On a hypersurface $\Sigma_{\alpha\beta\gamma}$ with constant $\alpha^{\pm}$, $\beta^{\pm}$, $\gamma$, the metric (\ref{btz}) reduces to
\begin{eqnarray}
ds^2 \Big|_{\Sigma_{\alpha\beta\gamma}}
&=&y^{-2} \, dy^2-y^{-4}(y^2 \cosh 2\gamma-4\beta^+\beta^-) dx^+dx^- \nonumber \\
&& +a(dx^+)^2+\bar{a}(dx^-)^2  \nonumber \\&&
+\frac{4}{y^4} \Big\{-a\beta^-(\beta^+)^3 -\bar{a} \beta^+(\beta^-)^3 \Big\} dx^+dx^- \nonumber \\
&& -\frac{a}{y^2} \Big \{ -3(\beta^+)^2\cosh 2\gamma +4\alpha^+ \sinh 2\gamma \Big\} dx^+dx^- \nonumber \\
&&- \frac{\bar{a}}{y^2} \Big\{ -3(\beta^-)^2\cosh 2\gamma +4\alpha^- \sinh 2\gamma \Big\} dx^+dx^- \nonumber \\
&& -a\bar{a}\Big\{ y^2 \cosh 2\gamma-6\beta^+\beta^--\frac{4}{y^4}(\beta^+\beta^-)^3 \nonumber \\
&& \qquad +\frac{1}{y^2} \{ 9(\beta^+\beta^-)^2 \cosh 2\gamma +16\alpha^+\alpha^- \cosh 2\gamma  \nonumber \\
&&\qquad -12(\beta^+)^2\alpha^-\sinh 2\gamma    
-12 (\beta^-)^2\alpha^+\sinh 2\gamma\} \Big\} dx^+dx^-. \nonumber \\
&&\label{BTZhyper} 
\end{eqnarray}
The induced metric (\ref{BTZhyper}) is also a solution to the equation of motion of 3d $SL(3,\mathbb{R}) \times SL(3,\mathbb{R})$ Chern-Simons gauge theory. So this is a new black hole solution in the 3d space-time. This black hole does not have spin-3 charge. On a hypersurface $\Sigma_0$ with $\alpha=\beta=\gamma=0$ this metric coincides with that of BTZ black hole\cite{BTZ}. This metric changes from one $\Sigma_{\alpha\beta\gamma}$ to another $\Sigma'_{\alpha\beta\gamma}$, when the values of $\beta^{\pm}$, $\alpha^{\pm}$, $\gamma$ are changed. If $\beta^{\pm} \neq 0$, the leading behavior of the metric near $y \sim 0$ is $y^{-4}$ and the  space-time is  asymptotically AdS with AdS length $=1/2$. 
As for the spin-3 field  we checked that as in (\ref{nabphi}) $\varphi$ for these flat connections ${\cal A}$, $\overline {{\cal A}}$ satisfy the 8d equation, 
\begin{equation}
\nabla_{\mu} \, \varphi_{\nu\lambda\rho}=0.  \label{metricBTZh}
\end{equation}
Result for the spin-3 field will not be presented here, because it is complicated.  
On the hypersurface $\Sigma_0$, where $\alpha^{\pm}=\beta^{\pm}=\gamma=0$, it vanishes.
\begin{equation}
\varphi|_{\Sigma_0} =0
\end{equation}
Hence the hypersuface $\Sigma_0$ is exactly the BTZ black hole. On other $\Sigma$'s spin-3 field $\varphi$ does not vanish.

The Hawking temperatures of the black hole (\ref{BTZhyper}) can be obtained by holonomy conditions\cite{CLM}\cite{Kraus}\cite{Lec}. Let us consider the case of finite and non-vanishing $a$ and $\bar{a}$. A matrix $U$  is defined by the flat connection ${\cal A}={\cal A}_0+\psi$ as 
\begin{equation}
{\cal A}= U^{-1} \, dU.
\end{equation}
On the hypersurface $\Sigma_{\alpha\beta\gamma}$ this reduces to ${\cal A}_{x^+}= U^{-1} \, \partial_{x^+} U$\footnote{Notice that ${\cal A}={\cal A}_{x^+} \, dx^+$ on the hypersurface and ${\cal A}_{x^+}$ does not depend on $x^+$.} and $U$ is solved as $U= \exp \big( x^+ \, {\cal A}_{x^+} \big)$.  
On the 3d Euclidean asymptotically AdS space, which is obtained by Wick rotation from $\Sigma$, the coordinates $x^+ =x+it_E\equiv z$ and $x^-=x-it_E\equiv -\bar{z}$ are 
identified as $(z, \bar{z}) \sim (z+2\pi \tau,\bar{z}+2\pi \bar{\tau})$, where $\tau$ and $\bar{\tau}$ are modular parameters of the boundary tori. 
A holonomy matrix $w$ is defined by $U(z,\bar{z})^{-1} \, U(z+2\pi \tau,\bar{z}+2\pi\bar{\tau})=\exp w$. 
Hence $w$ is given by 
\begin{equation}
w= 2\pi \tau \, {\cal A}_{x^+}. \label{wA}
\end{equation}
This is computed by using (\ref{AP1}),  (\ref{gau1}) and (\ref{psino1}). 
Similarly $\overline{{\cal A}}=\overline{{\cal A}}_0+\overline{\psi}$ defines $\bar{w}$. 
\begin{equation}
\bar{w}= 2\pi \bar{\tau} \, \overline{{\cal A}}_{x^-}.  \label{wbAb}
\end{equation}

By requiring that the flat connections are non-singular, the matrices $w$, $\bar{w}$ should be required to have the same eigenvalues as those for the vacuum. Hence they need to satisfy the conditions, \cite{Kraus}
\begin{eqnarray}
&& \text{det} \, w =0, \\
&& \text{tr} \, w^2=-8\pi^2,
\end{eqnarray}
and similar equations for $\bar{w}$. It can be shown that the first condition is trivially satisfied. The second one yields
\begin{equation}
\tau = \frac{i}{2\sqrt{a}}, \qquad \bar{\tau}= \frac{i}{2\sqrt{\bar{a}}} .
\end{equation}
Since $\tau$ is related to the inverse right and left  temperatures, $\beta_R$ and  $\beta_L$, as $\tau=\frac{i}{2\pi} \beta_R$ and $\bar{\tau}=\frac{i}{2\pi} \beta_L$, respectively, we obtain
\begin{equation}
T_R=\frac{1}{\beta_R}=\frac{\sqrt{a}}{\pi}= \frac{1}{\pi} \, \sqrt{2G(M+J)}. \label{rightT}
\end{equation}
where $M$ and $J$ are mass and angular momentum of the black hole. Similarly, for the left inverse temperature we have
\begin{equation}
T_L=\frac{1}{\beta_L}=\frac{\sqrt{\bar{a}}}{\pi}= \frac{1}{\pi} \, \sqrt{2G(M-J)}. \label{leftT}
\end{equation}
Hence $\alpha^{\pm}$, $\beta^{\pm}$ and $\gamma$ do not appear in the temperatures.

Now let us investigate whether the $\beta^+=-\beta^- \equiv \lambda \rightarrow \infty$ limit of the metric (\ref{BTZhyper}) exists.  Some calculation shows that 
even if coordinates $y$, $x^{\pm}$ are rescaled, \footnote{In the case of the metric (\ref{per1}) $y$ must also be rescaled as $y = \lambda^{1/2}\tilde{y }$ in order to take a finite limit of the metric as $\lambda \rightarrow \infty$, where $\beta^+=-\beta^-=\lambda$. } such a limit does not exist, unless $a=\bar{a}=0$. 
Hence in the case of a black hole, it is not possible to argue  existence of a fixed point along $\beta^+=-\beta^-\equiv\lambda$ by this method. This issue may be settled, if two-point functions of ${\cal W}$ primary operators in the boundary CFT at finite temperature, which is dual to (\ref{BTZhyper}), can be obtained. 

However,  if $a$ and $\bar{a}$ are also rescaled in an appropriate way,  finite limits exist. Because this may produce new solutions, we will study such limits.  We perform  the following rescaling of variables in (\ref{BTZhyper}),\footnote{Here the AdS length is set to 1. The metric on the hypersurface $\Sigma_{\alpha\beta\gamma}$ is a solution to the equations of motion of 3d $SL(3,\mathbb{R}) \times SL(3,\mathbb{R})$ Chern-Simons theory for each value of $\alpha^{\pm}$, $\beta^{\pm}$ and $\gamma$. On $\Sigma_{\alpha\beta\gamma}$, $\beta^{\pm}$ are not  coordinates, but just  constants. Hence the constants $a$, $\bar{a}$ and other variables can be rescaled and made dependent on $\lambda$.  } 
\begin{equation}
y =  \lambda^{\rho} \, \bm{y}, \qquad 
x^{\pm}  = \lambda^{\rho} \, \bm{x}^{\pm},  \label{temperatureflow1}
\end{equation}
as well as
\begin{equation}
a = \lambda^{-2\rho} \, \bm{a}, \qquad 
\bar{a}  \rightarrow  \lambda^{-2\rho} \, \bm{\bar{a}},  \label{temperatureflow2}
\end{equation}
while $\alpha^{\pm}$ and $\gamma$ are fixed.  The metric has a finite limit for the constant $\rho \geq 1$. The limit  depends on (1) $\rho=1$ or (2) $\rho >1$. 

\begin{enumerate}
\item [(1)] $\rho =1$:

In the  limit $\lambda \rightarrow \infty$, the 3d metric (\ref{BTZhyper}) asymptotes to 
\begin{eqnarray}
ds^2 \Big|_{\Sigma_{\alpha\beta\gamma}}
& \rightarrow &\bm{y}^{-2} \, d\bm{y}^2
 +\bm{a}(d\bm{x}^+)^2+\bm{\bar{a}}(d\bm{x}^-)^2  \nonumber \\&&
-\Big[\frac{4}{\bm{y}^4} (\bm{a} -1)(\bm{\bar{a}}-1)   +\frac{1}{\bm{y}^2} (3\bm{a}-1)(3\bm{\bar{a}}-1)\cosh 2\gamma   \nonumber \\
&& \quad +6\bm{a}\bm{\bar{a}} +\bm{y}^2\bm{a}\bm{\bar{a}}  \cosh 2\gamma \Big] d\bm{x}^+d\bm{x}^-. \label{BTZhyperasym} 
\end{eqnarray} 
If $(\bm{a}-1)(\bm{\bar{a}}-1) >0$, the signature of the metric is correct and this is an asymptotically AdS black hole solution with AdS length $\ell'_{\text{AdS}}=\frac{1}{2}$.  If $(\bm{a}-1)(\bm{\bar{a}}-1)= 0$, this is an asymptotically AdS black hole with the AdS length $\ell_{\text{AdS}}=1$. Metric (\ref{BTZhyperasym})  is a solution to the equations of motion in the spin-3 gravity based on the 3d Chern-Simons theory. 
The 3d metric  (\ref{BTZhyperasym}) depends on the parameter $\gamma$ in addition to $\bm{a}$ and $\bm{\bar{a}}$, the mass and angular momentum.
 As for the spin-3 field, it also has a well-defined $\lambda \rightarrow \infty$ limit.
\begin{eqnarray}
\varphi\Big|_{\Sigma_{\alpha\beta\gamma}} &\rightarrow& \frac{\tilde{a}}{y^2} \, dy^2dx^+-\frac{\bm{\bar{a}}}{y^2} \, dy^2dx^-\nonumber \\
&&+\frac{1}{2y^4}\Big[ (\bm{a}+1)^2(\bm{\bar{a}}-1)+\bm{a}\bm{\bar{a}}(3\bm{a}+1)y^4 \nonumber \\
&& \qquad -\bm{a}y^2\Big\{ (1+\bm{a})(1-3\bm{\bar{a}})-\bm{a}\bm{\bar{a}}y^4\Big\} \cosh 2\gamma \Big] (dx^+)^2dx^- \nonumber \\
&&+\frac{1}{2y^4}\Big[ (\bm{\bar{a}}+1)^2(-\bm{a}+1)-\bm{a}\bm{\bar{a}}(3\bm{\bar{a}}+1)y^4 \nonumber \\
&& \qquad +\bm{\bar{a}}y^2\Big\{ (1+\bm{\bar{a}})(1-3\bm{a})-\bm{a}\bm{\bar{a}}y^4\Big\} \cosh 2\gamma \Big] (dx^-)^2dx^+ \nonumber \\
&& -\frac{1}{2y^3}\Big[(1-3\bm{a})(1-3\bm{\bar{a}})-\bm{a}\bm{\bar{a}}y^4\Big] \, \sinh 2\gamma \, dy dx^+dx^-.  \label{BTZhyperasym1} 
\end{eqnarray}

\item [(2)] $\rho >1$:

In the  limit $\lambda \rightarrow \infty$, the 3d metric (\ref{BTZhyper}) asymptotes to 
\begin{eqnarray}
ds^2 \Big|_{\Sigma_{\alpha\beta\gamma}}
& \rightarrow &\bm{y}^{-2} \, d\bm{y}^2
 +\bm{a}(d\bm{x}^+)^2+\bm{\bar{a}}(d\bm{x}^-)^2  
-\cosh 2\gamma \Big[\bm{y}^{-2}  +\bm{a}\bm{\bar{a}}\bm{y}^2  \Big]d\bm{x}^+d\bm{x}^-.
 \label{BTZhyperasym2} 
\end{eqnarray}
This coincides with the metric at $\alpha^{\pm}=\beta^{\pm}=0$
There are terms which contain $\gamma$ in (\ref{BTZhyperasym2}). Due to the factor $\cosh \gamma$ this is a deformed BTZ solution. 
As for the spin-3 field, it also has a well-defined $\lambda \rightarrow \infty$ limit.
\begin{equation}
\varphi\Big|_{\Sigma_{\alpha\beta\gamma}} \rightarrow \frac{1}{2\bm{y}^3} \, (-1+\bm{a}\bm{\bar{a}} \bm{y}^4) \sinh 2\gamma \, d\bm{x}^+ d\bm{x}^-d\bm{y}
\end{equation}
\end{enumerate}

\subsection{Black Hole Solution with Spin-3 Charges}
\hspace{5mm}
In this subsection the functions $Q_n$, $\overline{Q}_n$ and connections $\psi$, $\overline{\psi}$ for the black hole solution with spin-3 charge will be constructed. This is more difficult than the preceding black hole, because more parameters than the mass and angular momentum must be introduced and the integrability condition for the partition function needs to be taken into account\cite{Kraus}. The result is presented in (\ref{Q-1}), (\ref{psi1}) in appendix G. The result includes parameters $b, \bar{b}, \mu, \bar{\mu}$,  which are spin-3 charges and chemical potentials, in addition to $a$ and $\bar{a}$.  $\psi^{(1)}$ and  $\overline{\psi}^{(1)}$ are  contributions to  first order of these parameters.  When the black hole has spin-3 charges, however, $\psi$ and  $\overline{\psi}$ do not stop at the first order. It is then necessary to expand $\psi$ as $\psi=\psi^{(1)}+\psi^{(2)}+\cdots$ and systematically solve the equation for $\psi^{(i)}$.   The $i$-th-order perturbation $\psi^{(i)}$ needs to satisfy
\begin{equation}
d\psi^{(i)}+{\cal A}_0 \wedge \psi^{(i)}+\psi^{(i)} \wedge {\cal A}_0 
=-\sum_{k=1}^{i-1} \psi^{(k)} \wedge \psi^{(i-k)}.  \label{psi-i}
\end{equation}
Suppose that $\psi^{(k)}$ for $k=1,2, \cdots, i-1$ has been obtained. Then the right hand side of (\ref{psi-i}) can be evaluated by using these results. By taking appropriate linear combinations of the  $a$-th components ($a=1,2, \cdots , 8$ ) of the left hand side such that the left hand side becomes exact  forms, the equations for $\psi^{(i)a}$ can be solved locally. Ambiguities associated with the solutions to the homogeneous equations for $\psi^{(i)a}$ must be used to make $\psi^{(i)a}$ periodic in $x^+$ and to keep Fefferman-Graham gauge for the metric, {\em i.e.}, $\psi^{(i)2}=0$. This forces us to introduce new functions $Q^{(2)}_n$ and $Q^{(3)}_n$ at the second and third orders of perturbation, which will yield new contributions to $\psi^{(2)a}$ and $\psi^{(3)a}$ according to (\ref{psia}).  It is also necessary to introduce other new terms to $\psi^{(2)a}$, which are proportional to the second order infinitesimal parameters, $\mu a$, $\mu b$, but otherwise must have been included in the first order connection $\psi^{(1)a}$.  The results for $\psi^{(2)}$ and $\overline{\psi}^{(2)}$ are given in (\ref{psi2}). We carried out analysis to the fourth order and  the results  for $\psi^{(3)}$ are presented in (\ref{psi3}). These solutions contain extra constant parameters $\zeta_i$, $\bar{\zeta}_i$ ($i=1,2,3$), which will be determined by the condition of integrability shortly. As will be clear from the result, the solution $\psi$ does not have -- components, while $\overline{\psi}$ does not have + components.  The perturbation expansions do not seem to terminate at a finite order. 

The metric of the black hole solution with spin-3 charge up to the first order in $b, \bar{b}, \mu, \bar{\mu}$  is given by 
\begin{equation}
ds^2 = ds_0^2+ds_1^2, \label{ds2withcharge}
\end{equation}
where $ds^2_0$ is given in (\ref{8dmetric}) and the other terms are presented in (\ref{ds2withcharge}). On $\Sigma_0$ the black hole metric does not coincide with that of the black hole solution with spin-3 charge obtained in \cite{Kraus}. 

Let us consider holonomy properties of this black hole. 
In this case the non-vanishing element of  ${\cal A}_{\mu}$ on $\Sigma_{\alpha\beta\gamma}$ is only ${\cal A}_{x^+}$. 
Then ${\cal A}= U^{-1}dU$ is solved as $U=\exp (x^+{\cal A}_{x^+})$ and the holonomy is given  as (\ref{wA}).
Then up to the fourth order of perturbation the conditions (1) $\text{det}\,  w=0$ and (2) $\text{tr}\, w^2+8\pi=0$ are given by the equations,
\begin{eqnarray}
(1)&&27b^2 \zeta_2^2\mu^3\tau^3+9b\tau^3 \Big\{3+a(-6\zeta_2+\zeta_1(8+\zeta_2))\mu^2\Big\}  \nonumber \\
&& \ +a^2\mu\tau^3 \Big\{27+27\zeta_3-2a\zeta_1^3\mu^2
+9\zeta_1(-1+8a\zeta_3\mu^2)\Big\} =\text{(4th-order terms)}, \label{detb} \\
(2)&&3+12a\tau^2-12b(-4+\zeta_2)\mu\tau^2+4a^2(\zeta_1^2+12\zeta_3)\mu^2\tau^2= \text{(5th-order terms)}, \label{Trb}
\end{eqnarray}
respectively. Here $\zeta_1$, $\zeta_2$ and $\zeta_3$ are constants like $a$, $b$, $\mu$ in the solutions as mentioned above. On the right hand sides of the above equations there are terms which are fourth- and fifth-order in the expansion parameters, respectively, and depend also on $\alpha^+$ and $\beta^+$. At the present (fourth) order of perturbation some terms on the left hand side of the first equation are of higher orders than those on the right hand side. The left hand sides do not depend on the values of $\alpha^+$, $\beta^+$ and $\gamma$ of $\Sigma_{\alpha\beta\gamma}$. We found that at each order of the perturbation theory (from the first to the fourth order)  the leading terms on the right hand sides are 
 exactly canceled out by new contributions from the next-order perturbation and instead, new terms which are higher orders in the expansion parameters  appear. Those terms on the left hand sides in the above two equations still do not change. If this pattern observed in the perturbative calculation persists to all higher orders, ultimately (\ref{detb}) and (\ref{Trb})  will converge to the following. 
\begin{eqnarray}
(1)&&27b^2 \zeta_2^2\mu^3\tau^3+9b\tau^3 \Big\{3+a(-6\zeta_2+\zeta_1(8+\zeta_2))\mu^2\Big\}  \nonumber \\
&& \qquad +a^2\mu\tau^3 \Big\{27+27\zeta_3-2a\zeta_1^3\mu^2
+9\zeta_1(-1+8a\zeta_3\mu^2)\Big\} =0, \label{det} \\
(2)&&3+12a\tau^2-12b(-4+\zeta_2)\mu\tau^2+4a^2(\zeta_1^2+12\zeta_3)\mu^2\tau^2=0 \label{Tr}
\end{eqnarray}
Henceforth we will assume that this is valid.  

Then  (\ref{Tr}) is solved as 
\begin{eqnarray}
b=\frac{1+4\tau^2 a+(1/12)(\zeta_1^2+12\zeta_3)\nu^2a^2}{(\zeta_2-4)\nu \tau}.
\end{eqnarray}
Here $\nu$ is defined by
\begin{equation}
\nu = 4\mu \tau.  \label{nu}
\end{equation}
By substituting this into (\ref{det}) and differentiating the result with respect to $\tau$ and $\nu$, respectively, and we solve the differentiated equations for $(\partial a/\partial \nu)_{\tau}$ and $(\partial b/\partial \tau)_{\nu}$.  Further we require that these two differential coefficients should satisfy
\begin{equation}
\Big(\frac{\partial a}{\partial \nu}\Big)_{\tau}=\,  \Big( \frac{\partial b}{\partial \tau }\Big)_{\nu}. \label{integrability}
\end{equation}
This equation (\ref{integrability}) is the integrability condition for the partition function of the black hole\cite{Kraus},
\begin{equation}
Z = \text{Tr} \, e^{\frac{i\pi}{2G}\tau (a +\mu b)} \, e^{-\frac{i\pi}{2G}\bar{\tau}(\bar{a}+\bar{\mu} \bar{b})}.  
\end{equation}
$U=a/4G$ is the energy, $q=-b/G$  the spin-3 charge and $\mu$ the chemical potential. 
These conditions determine $\zeta_n$  ($n=1,2,3$) uniquely.
\begin{equation}
\zeta_1=-2, \quad \zeta_2=-8, \quad \zeta_3=1 \label{zetas}
\end{equation}
Then the right-mover  temperature is given by $T_R= i/(2\pi \tau)$. $\tau$ and $\nu (=4\mu \tau)$ can be determined from (\ref{det}) and (\ref{Tr}) as  functions of 
$a$ and $b$. A similar analysis for the left mover can also be carried out. The conditions for $\bar{w}$ similar to (\ref{det}) and (\ref{Tr}) with bars on the parameters ensure the integrability of the partition function, if the following values of the parameters are chosen.
\begin{equation}
\bar{\zeta}_1=-2, \quad \bar{\zeta}_2=8, \quad \bar{\zeta}_3=1 \label{zetabars}
\end{equation}
We need to set $\bar{\nu}=4\bar{\mu}\bar{\tau}$. Then the left-mover  temperature is given by $T_L=- i/(2\pi \bar{\tau})$. Entropy of the black hole $S=S_R+S_L$  can be obtained by the method used in \cite{Kraus}. The right-moving part $S_R$ is given by $S_R= (\pi\ell_{\text{AdS}}/2G) \sqrt{a}f(27b^2/2a^3)$, where  $f(y) = \cos \theta$, $\theta = \arctan [\sqrt{y(2-y)}/6(1-y)]$. The entropy and partition function do not depend on $\alpha$ and $\beta$. 
A scalar field operator in this black hole background will be dual to all ${\cal W}$-descendants of some scalar operator in a ${\cal W}_3$-extended CFT at finite temperature.

\subsection{Black Hole Solution with Spin-3 Charge on $\Sigma_0$}
\hspace{5mm}
As mentioned above, although the flat connections  are already complicated even at the fourth order of perturbation, miraculous cancellation occurs in the holonomy conditions, (\ref{det}) and (\ref{Tr}). The variables $\alpha^{\pm}$, $\beta^{\pm}$ and $\gamma$ corresponding to  $\Sigma_{\alpha\beta\gamma}$ do not appear in these conditions. This situation is similar to that in the black hole solution without spin-3 charge, which was observed in subsec. 5.1. So, let us study  the holonomy conditions for the flat connections on the hypersurface $\Sigma_0$, where $\alpha^{\pm}=\beta^{\pm}=\gamma=0$. From the results in appendix D,  after setting $\alpha^{\pm}=\beta^{\pm}=\gamma=0$  the flat connections on $\Sigma_0$ are given by
\begin{eqnarray}
{\cal A}|_{\Sigma_0} &=& \Big((-a+\zeta_2 b \mu)t_1+t_3-(b+\zeta_3\, a^2\mu)t_4-\zeta_1a \mu \, t_6 -\mu t_8\Big) \, dx^+, \label{f1}\\
\overline{{\cal A}}|_{\Sigma_0} &=& \Big( -t_1+(\bar{a}+\bar{\zeta}_2\bar{b}\bar{\mu})\,t_3-\bar{\mu} t_4-\bar{\zeta}_1\,\bar{a}\bar{\mu}\, t_6-(\bar{b}+\bar{\zeta}_3 \,\bar{a}^2\bar{\mu}) \, t_8  \Big) \, dx^-. \label{f2}
\end{eqnarray}
These are solutions to the equations of motion in 3d Chern-Simons theory with the  boundary conditions, ${\cal A}_{-}=0$ and $\overline{{\cal A}}_{+}=0$. 
Because these connections are proportional to $dx^+$ and $dx^-$, respectively, the flatness condition is trivially satisfied and  they do not determine the parameters.  Only the conditions of holonomy (\ref{det})  and (\ref{Tr}) determine these parameters and the results agree with (\ref{zetas}) and (\ref{zetabars}). These are the only consistent conditions. Hence this may support  the expectation that the conditions  (\ref{det}) and (\ref{Tr}) will remain valid, even if higher-order terms which depend on $\alpha$, $\beta$, $\gamma$ are included in $\psi$ and $\overline{\psi}$. 
From (\ref{f1})-(\ref{f2}) the metric is obtained as follows. 
\begin{eqnarray}
ds^2|_{\Sigma_0} &=& \frac{1}{y^2} dy^2+\big( a+12b\mu+\frac{16}{3}a^2\mu^2\big)(dx^+)^2+\big( \bar{a}+12\bar{b}\bar{\mu}+\frac{16}{3}\bar{a}^2\bar{\mu}^2\big) (dx^-)^2 \nonumber \\
&& -\frac{1}{3y^4}\big\{ 3y^2+12\mu\bar{\mu}+8a\bar{a}\mu\bar{\mu}y^4+12y^8(b+a^2\mu)(\bar{b}+\bar{a}^2\bar{\mu})  \nonumber \\
&&+3y^6(a+8b\mu)(\bar{a}+8\bar{b}\bar{\mu})\big\} dx^+dx^- \label{metricSigma0}
\end{eqnarray}

General black hole solutions with spin-3 charge will be obtained by making $a$, $b$ and $\mu$ in ${\cal A}|_{\Sigma_0}$ depend on $x^+$. Similarly, $\bar{a}$, $\bar{b}$ and $\bar{\mu}$ in  $\overline{{\cal A}}|_{\Sigma_0} $ are replaced by functions of $x^-$.
\begin{eqnarray}
{\cal A}|_{\Sigma_0} &=& \Big\{-\big(a(x^+)+8 b(x^+) \mu(x^+)\big)t_1+t_3-\big(b(x^+)+ a(x^+)^2\mu(x^+)\big)t_4 \nonumber \\
&&+2a(x^+) \mu(x^+) \, t_6 -\mu(x^+) t_8\Big\} \, dx^+, \nonumber \\
\overline{{\cal A}}|_{\Sigma_0} &=& \Big\{ -t_1+\big(\bar{a}(x^-)+8\bar{b}(x^-)\bar{\mu}(x^-)\big)\,t_3-\bar{\mu}(x^-) t_4+2\,\bar{a}(x^-)\bar{\mu}(x^-)\, t_6 \nonumber \\
&&-\big(\bar{b}(x^-)+\bar{a}(x^-)^2\bar{\mu}(x^-)\big) \, t_8  \Big\} \, dx^-. \label{ff2}
\end{eqnarray}
These connections are still flat. These connections are further required to satisfy the holonomy conditions $\text{det} \, w=0$ and $\text{tr} \, w^2=-8\pi^2$, which are similar to (\ref{det}) and (\ref{Tr}) but more complicated. Here the holonomy matrix $w$ is defined as before now by using a path-ordered exponential $U(x^+)=P \exp \{ -\int^{x^+}_{x^+_0}{\cal A}_{+}(x'^+)dx'^+\}$.   It is known that when $b=\mu=\bar{b}=\bar{\mu}=0$, the metric constructed from these connections is the most general BTZ metric in the Fefferman-Graham gauge\cite{BTZ}. When ${\cal A}$ is transformed  as ${\cal A} \rightarrow U^{-1} {\cal A}U+U^{-1}dU$ with $U=\exp t_a \lambda^a(x^+)$ by restricting infinitesimal parameters $\lambda^a$ to  keep the form of ${\cal A}$ and imposing $\delta \mu=0$, transformations  $\delta a$ and $\delta b$ are obtained. For $\mu=0$  these transformations generate the ${\cal W}_3$ algebra.\cite{Campoleoni} 

Next we consider connections which do not satisfy $A_{-}, \overline{A}_{+} = 0$. 
\begin{eqnarray}
{\cal A}' &=& \big( t_3-at_1-bt_4\big) dx^++\mu \big( 8b t_1+a^2 t_4-2at_6+t_8 \big)dx^-, \label{g1}\\
\overline{{\cal A}} \,'  &=& -\big(t_1-\bar{a}t_3-\bar{b} t_8\big) dx^-+\bar{\mu} \big(8\bar{b}t_3+t_4-2\bar{a}t_6+\bar{a}^2t_8 \big) dx^+ \label{g2}
\end{eqnarray}
It can be shown that these connections are flat and up to change of notations these coincide with eq (4.1) of \cite{BH}. 
Then the metric derived from these connections is given as follows and does not coincide with (\ref{metricSigma0}). 
\begin{eqnarray}
ds'^2 &=& \frac{1}{y^2}dy^2+\frac{1}{3}\big\{ 3a(1+8\bar{b}\bar{\mu} y^2)+4\bar{a}^2\bar{\mu}(3by^4+4\bar{\mu})\big\}(dx^+)^2 \nonumber \\
 &&+\frac{1}{3}\big\{ 3\bar{a}(1+8b\mu y^2)+4a^2\mu(3\bar{b}y^4+4\mu)\big\} (dx^-)^2 \nonumber \\
&& -\Big\{ \frac{1}{y^2}+\frac{4\mu\bar{\mu}}{y^4}+\frac{4}{3}(9b\mu+9\bar{b}\bar{\mu}+2a\bar{a}\mu\bar{\mu}) \nonumber \\
&& +4y^4(b\bar{b}+a^2\bar{a}^2\mu\bar{\mu})+y^2(a\bar{a}+64b\bar{b}\mu\bar{\mu})\Big\}dx^+dx^-  \label{ds'2}
\end{eqnarray}
It is checked that the spin-3 field obtained from (\ref{f1})-(\ref{f2}) are also different from that computed from (\ref{g1})-(\ref{g2}). The conditions of holonomy can be obtained 
by $w=2\pi (\tau {\cal A}'_+-\bar{\tau} {\cal A}'_-)$ and its barred counterpart. The conditions (1) $\text{det}\,  w=0$ and (2) $\text{tr}\, w^2+8\pi=0$ read
\begin{eqnarray}
(1)&&27b^2 \nu^3 +18a^2\nu \tau^2-2a^3\nu^3+27ab\nu^2\tau+27b\tau^3=0, \label{det'} \\
(2)&&3+12a\tau^2+36b \nu \tau+4a^2\nu^2=0, \label{Tr'}
\end{eqnarray}
where $\nu=4\bar{\tau}\mu$. These exactly coincide with (\ref{det})-(\ref{Tr}) when (\ref{nu}) and (\ref{zetas}) are substituted. Similar conditions for the barred quantities also coincide.\footnote{For barred quantities we must set $\bar{\nu}=4\tau \bar{\mu}$.}
Then it can be checked that the integrability condition (\ref{integrability}) also holds in the gravity theory defined by the flat connections (\ref{g1}) and (\ref{g2}). Hence although the asymptotic behaviors of (\ref{f1})-(\ref{f2}) and  (\ref{g1})-(\ref{g2}) as $y \rightarrow 0$ are different, and the 3d flat connections are not gauge equivalent, these two sets of flat connections define two distinct bulk gravity geometries which have the same partition function, when the parameters  are appropriately identified.  This means that the partition functions of the would-be CFT's dual to each backgrounds will coincide.  These 8d flat connections  might be related by large gauge transformations. 

Let us study the boundary conditions for the connections. The variation of the CS action (\ref{CS}) is given by 
\begin{equation}
\delta S_{\text{CS}} [A']=\frac{k}{2\pi}\int_M \text{tr} \, \delta A' \wedge (dA'+A' \wedge A')+\frac{k}{4\pi}\int_{\partial M} \text{tr}\,  (A'_{+} \delta A'_{-}-A'_{-} \delta A'_{+})d^2x.
\end{equation}
To make this vanish after the bulk equation of motion is used,  $\delta A'$ must satisfy 
$\text{tr}\,  (A'_{+} \delta A'_{-}-A'_{-} \delta A'_{+})=0 $ on the boundary.
Usually, the boundary condition $A'^a_{-}=0$ or $A'^a_{+}=0$ is imposed as (\ref{ff2}). 
The connection (\ref{g1}), however, satisfies the following conditions, one for each canonical pair, 
\begin{eqnarray}
A'^3_+=y^{-1}, \quad A'^8_-=\mu y^{-2}, \quad 
A'^2_-=A'^3_-= A'^7_-=A'^6_+=A'^7_+=A'^8_+=0, 
\end{eqnarray}
where $A'$ is obtained from ${\cal A}'$ by the transformation (\ref{gauge1}). These determine the Dirichlet conditions. Then a  variation of the action vanishes after addition of extra local terms $(k/4\pi)\int_{y=\epsilon} \, (4y^{-1} \, A'^1_-+16\mu \,  y^{-2}A'^4_+\, ) \, d^2x$ to the action (\ref{CS}). Here $\epsilon$ is a UV cutoff and finally a limit $\epsilon \rightarrow 0$ must be taken. Under this variation $\mu$ should not be changed. Hence there exist appropriate boundary conditions.

Remaining problem is how to compute partition functions in an  explicit manner. 
It can also be shown that the metric (\ref{metricSigma0}) is in a wormhole gauge as the black hole solution in \cite{Kraus}. Note, however,  that in the limit $a$, $b$, $\mu \rightarrow 0$ only the flat connections (\ref{f1})-(\ref{f2}) correspond to the 8d vielbein which reproduces the spin-3 field (\ref{phiCubic}) and the coefficient function of the cubic equation for the scalar field.

\section{Summary and Discussions}
\hspace{5mm}
In this paper a formulation of 3d spin-3 gravity coupled to a scalar field is studied from the view point of a realization of ${\cal W}_3$ symmetry in the bulk space-time, not in the local frame. It is shown that this formulation is possible in the extended 8d space. In the most symmetric case this is a group manifold $SU(1,2)$. In this 8d space holographic duality between the bulk and the boundary is explicitly realized. The ordinary 3d bulk is obtained by restricting the space-time to hypersurfaces $\Sigma_{\alpha\beta\gamma}$ which have constant values of $\alpha^{\pm}$, $\beta^{\pm}$ and $\gamma$.  The action for both spin-3 gravity and matter fields are obtained explicitly as 8d integrals. To the leading order of $1/c$ expansion the 8d vielbein field $e^a_{\mu}$ is obtained as a solution to the problem of flat connections in 8d space-time, and new black hole solutions are obtained. In this case the 8d space is a deformation of the $SU(1,2)$ manifold.     The black hole solutions with and  without spin-3 charge are found and their flat connections satisfy $A_{x^-}=\overline{A}_{x^+}=0$.  The partition function of the black hole with spin-3 charge on $\Sigma_0$ is found to coincide with that of the black hole solution with different boundary condition obtained before in \cite{Kraus}. Further investigation of the black hole solution (\ref{f1})- (\ref{f2}) is necessary. 

In sec. 1 it is shown that the scaling dimension $\Delta$ of a scalar operator on the boundary  is related to the scalar mass $m$ and spin-3 charge $q$ by (\ref{Deltamass}). This is similar to the ordinary dictionary for the simple AdS$_3$ gravity, $\Delta=1+\sqrt{1+m^2}$, but slightly modified. 
In the case of 3d higher-spin gauge theory dual to ${\cal W}_N$ minimal model it was shown that a scalar field can be consistently coupled to $hs[\mu]$ theory, if and only if the mass of the scalar field satisfies $m^2=\mu^2-1$.\cite{PV}\cite{GG}  
In the case of the spin-3 theory it turns out $\mu=N=3$ and hence $m^2=8$ is singled out.  
The mass formula (\ref{Deltamass}), however, does not restrict the value of $m$. This is not a contradiction, because the two models of spin-3 gravity are distinct. Let us note that the central charge $c$ of the algebra (\ref{nonlinerW3}) is arbitrary in principle, and it is taken to be  infinite in the semi-classical treatment, while the central charge (\ref{cNk}) for $N=3$ is finite. The scalar field $\Phi$ in our model has W$_0$ charge $q$.

There are several questions which are not considered in this paper.  In the case of the bulk space-time  with the full $su(1,2) \times su(1,2)$ symmetry, a scalar field satisfies the cubic-order differential equation (\ref{thirdorder})  in addition to Klein-Gordon equation (\ref{KG}).  
Then, in the case of  black hole solutions, or in the case of more general asymptotically AdS space-times, does a cubic-order differential equation, which is compatible with Klein-Gordon equation, exist and  is it related to the spin-3 gauge field as in (\ref{thirdorder})? 
When there is no matter field, the $SL(2,\mathbb{R}) \times SL(2,\mathbb{R})$ Chern-Simons theory of 3d gravity is renormalizable\cite{W}. When there is no matter field in our 8d formulation of spin-3 gravity, the classical equations of motion are also conditions of flat connections and there is no physical degrees of freedom inside the bulk. When the action integral is rewritten into the metric-like formalism, how is the absence of  graviton in the bulk ensured?  In the case of the ordinary 3d gravity Riemann tensor can be expressed in terms of Ricci tensor. Is there a similar identity in spin-3 gravity?   
In this paper a bulk-to-boundary propagator for a scalar field is calculated. If the bulk-to-bulk propagator is obtained, conformal blocks of ${\cal W}_3$ extended CFT may be studied by using the methods of holography. 

Finally, we had to introduce 8d space-time, which is a deformation of $SU(1,2)$, to realize holographic duality of ${\cal W}_3$ CFT and spin-3 gravity. In \cite{W1}\cite{W2}\cite{Sund}\cite{Mari} it was argued that the higher spin gauge theory might appear as some special limiting case  of string field theory. Recently, there are also works on higher-spin gauge theory from the point of view of string theory and the tensionless limit is studied.\cite{GGH}\cite{FGJ}\cite{GG2}\cite{EGG}  These are related to ${\cal W}_{\infty}$ algebra. 
It is not clear how to embed the result of the present paper in the string theory.  
 We hope to report on this issue  elsewhere.

\setcounter{section}{0}
\renewcommand{\thesection}{\Alph{section}}
\section{Conventions}
\hspace{5mm}
Conventions for generators of $sl(3,\mathbb{R})$ and $su(1,2)$ algebras are summarized. 
\subsection{ $sl(3,\mathbb{R})$ Algebra}
\hspace{5mm}
 Generators of $s\ell(3,R)$ in the fundamental representation are given \cite{Campoleoni} by
\begin{eqnarray}
t_1 &=& \left(\begin{array}{ccc}
        0  & 0& 0 \\
        1 & 0 & 0 \\
        0 & 1 & 0  \end{array}\right), \qquad 
t_2  = \left(\begin{array}{ccc}
        1  & 0& 0 \\
        0 & 0 & 0 \\
        0 & 0 & -1 \end{array} \right), \qquad 
t_{3}  = \left(\begin{array}{ccc}
        0  & -2& 0 \\
        0 & 0 & -2 \\
        0 & 0 & 0  \end{array}\right), \nonumber \\
t_4 &=& \left(\begin{array}{ccc}
        0  & 0& 0 \\
        0 & 0 & 0 \\
        2 & 0 & 0  \end{array}\right), \qquad 
t_5  = \left(\begin{array}{ccc}
        0  & 0& 0 \\
        1 & 0 & 0 \\
        0 & -1 & 0 \end{array} \right), \qquad 
t_6  = \frac{2}{3} \, \left(\begin{array}{ccc}
        1  & 0& 0 \\
        0 & -2 & 0 \\
        0 & 0 & 1  \end{array}\right), \nonumber \\
t_{7} &=& \left(\begin{array}{ccc}
        0  & -2 & 0 \\
        0 & 0 & 2 \\
        0 & 0 & 0 \end{array} \right), \qquad 
t_{8}  = \left(\begin{array}{ccc}
        0  & 0& 8 \\
        0 & 0 & 0 \\
        0 & 0 & 0  \end{array}\right)
\label{generators}
\end{eqnarray}
The structure constants ${f_{ab}}^c$ are defined by 
\begin{eqnarray}
 [ t_a, t_b]= {f_{ab}}^c \, t_c.
\end{eqnarray}
and a Killing metric $h_{ab}$ is given by 
\begin{eqnarray}
h_{ab}=\frac{1}{2} \, \mbox{tr} \, (t_a t_b), 
\end{eqnarray}
and its nonzero components are  $h_{22}=1, \ h_{13}=-2, 
h_{48}=8, \ h_{57}=-2, \ h_{66}=4/3$. 
Indices of the local frame are raised and lowered by $h_{ab}$ and its inverse 
$h^{ab}$. Then $f_{abc} \equiv {f_{ab}}^d \, h_{dc}$ 
is completely anti-symmetric. Non-vanishing structure constants are given by
\begin{eqnarray}
&& f_{123} = -2, \ f_{158}=8, \ f_{167}=-4, \ f_{248}=-16, \nonumber \\
&& f_{257} = 2, \ f_{347}=8, \ f_{356}=-4
\end{eqnarray}

The completely symmetric invariant tensor $d_{abc}$ is defined by 
\begin{eqnarray}
d_{abc}=\frac{1}{2} \, \text{tr} \{t_a, t_b \} t_c,
\end{eqnarray}
These constants are given by
\begin{eqnarray}
&& d_{127}=d_{235}=-2, \quad  d_{136}= d_{226}=d_{567}=\frac{4}{3},  \quad 
d_{118}= d_{334}=8, \nonumber \\
&& d_{468}=\frac{32}{3}, \quad d_{477}= d_{558}=-8, \quad  d_{666}=-\frac{16}{9}.
\end{eqnarray}

Casimir operators are given by 
\begin{eqnarray}
C_2 &=& h^{ab} \, T_a \, T_b, \\
C_3 &=& d^{abc} \, T_a \, T_b \, T_c,
\end{eqnarray}
where $T_a$ is some irreducible representation of $sl(3,\mathbb{R})$. For the adjoint representation ${(T^a)^b}_c=-{{f_b}^a}_c$
 the following relations are obtained. 
\begin{eqnarray}
 {f_a}^{cd} \, f_{bcd} &=& -12 h_{ab}, \label{hff} \\
d^{\, cde} \, f_{caf}{f_d\,}^{fg}f_{egb} &=& 0
\end{eqnarray}

\subsection{$su(1,2)$ Algebra}
\hspace{5mm}
Generators of $su(1,2)$ algebra are obtained by replacing the matrices  of $sl(3,\mathbb{R})$ generators (\ref{generators}) as $t_a \rightarrow \tilde{t}_a$, where 
\begin{eqnarray}
\tilde{t}_j &=& t_j \quad (j=1,2,3), \qquad 
\tilde{t}_{\alpha} = i \, t_{\alpha} \quad (\alpha=4, \ldots 8).
\end{eqnarray}
These new matrices satisfy the relation $(\tilde{t}_a)^{\dagger} \, \eta +\eta \, \tilde{t}_a =0$, where dagger stands for  hermitian conjugation and $\eta$ is the following matrix.
\begin{eqnarray}
\eta &=& \left(\begin{array}{ccc}
        0  & 0& 1 \\
        0 & -1 & 0 \\
        1 & 0 & 0  \end{array}\right)
\end{eqnarray}
The matrices $\tilde{t}_a$  generate the ${\cal W}_3$ wedge algebra (\ref{W3wedgealgebra}) after an identification 
 $\tilde{t}_1=L_1$, $\tilde{t}_2=L_0$, $\tilde{t}_3=L_{-1}$, $\tilde{t}_4=W_2$, $\tilde{t}_5=W_1$, $\tilde{t}_6=W_0$, $\tilde{t}_7=W_{-1}$ and $\tilde{t}_8= W_{-2}$,  

Killing metric $\tilde{h}_{ab}=(1/2) \, \text{tr} \,(\tilde{t}_a \, \tilde{t}_b)$ is given by $\tilde{h}_{22}=1$, $\tilde{h}_{13}=-2$, $\tilde{h}_{48}=-8$, $\tilde{h}_{57}=2$, $\tilde{h}_{66}=-4/3$. The structure constants are given by $\tilde{f}_{123} = -2$, $\tilde{f}_{158}=-8$,  $\tilde{f}_{167}=-4$,  $\tilde{f}_{248}=16$, 
$\tilde{f}_{257} =-2 $, $\tilde{f}_{347}=-8$, $\tilde{f}_{356}=4$. The constants $\tilde{d}_{abc}$  are given by 
$\tilde{d}_{127}=\tilde{d}_{235}=-2i$, $\tilde{d}_{136}= \tilde{d}_{226}=-\tilde{d}_{567}=\frac{4}{3}i$,   
$\tilde{d}_{118}= \tilde{d}_{334}=8i$, $\tilde{d}_{468}=-\frac{32}{3}i$, $ \tilde{d}_{477}= \tilde{d}_{558}=8i$,  $\tilde{d}_{666}=\frac{16}{9}i$.
Casimir operators are given by 
\begin{eqnarray}
\tilde{C}_2 &=& \tilde{h}^{ab} \, \tilde{t}_a \, \tilde{t}_b, \\ \label{C2tilde}
\tilde{C}_3 &=& \tilde{d}^{abc} \, \tilde{t}_a \, \tilde{t}_b \, \tilde{t}_c. \label{C3tilde}
\end{eqnarray}

\section{Representation of ${\cal W}_3$ Generators by Differential Operators}
\hspace{5mm}
Here the infinite-dimensional representation of $su(1,2) \times su(1,2)$ algebra in the hyperbolic representation\cite{GT}  is presented. 
This is a representation for the generators of transformations in the {\em bulk}.
This is obtained from eqs (4.3) and (4.4) of \cite{NS} by replacements $\alpha^{\pm} \rightarrow i\alpha^{\pm}$, $\beta^{\pm} \rightarrow i\beta^{\pm}$, 
$\gamma \rightarrow i\gamma$.
 \begin{eqnarray}
L_{-1}^h &=& i\partial_{x^+}, \nonumber \\
L_0^h &=& -x^+\partial_{x^+}-2\alpha^+\partial_{\alpha^+}-\beta^+\partial_{\beta^+}-\frac{1}{2}y\partial_y, \nonumber \\
L_1^h &=& -i[(x^+)^2+3(\beta^+)^2]\partial_{x^+} -ix^+y\partial_y-3i\beta^+\partial_{\gamma}-i[2(\beta^+)^3+4x^+\alpha^+]\partial_{\alpha^+}
\nonumber \\
&&-i[2x^+\beta^++4\alpha^+]\partial_{\beta^+}-iy^2 \cosh (2\gamma) \partial_{x^-}-iy^2\cosh (2\gamma)\beta^-\partial_{\alpha^-}-iy^2\sinh (2\gamma)\partial_{\beta^-}, \nonumber \\
&& \label{diffL}
\end{eqnarray}
\begin{eqnarray}
W_{-2}^h &=& i\partial_{\alpha^+}, \nonumber \\
W_{-1}^h &=& -x^+\partial_{\alpha^+}-\partial_{\beta^+}, \nonumber \\
W_0^h &=& -2i\beta^+\partial_{x^+}-i [(x^+)^2+(\beta^+)^2]\partial_{\alpha^+}-2ix^+\partial_{\beta^+}-i\partial_{\gamma}, \nonumber \\
W_1^h &=& 3x^+\partial_{\gamma}+[-4\alpha^++6x^+\beta^+] \partial_{x^+}+[(x^+)^3+3x^+(\beta^+)^2]\partial_{\alpha^+}+[3(x^+)^2+(\beta^+)^2] \partial_{\beta^+} \nonumber \\
&& +\beta^+y\partial_y+y^2 \sinh (2\gamma)\partial_{x^-}+\beta^-y^2\sinh(2\gamma) \partial_{\alpha^-}+y^2 \cosh (2\gamma)\partial_{\beta^-}, \nonumber \\
W_2^h &=& -i \big[3(\beta^+)^4-(x^+)^4+16(\alpha^+)^2-6(x^+)^2(\beta^+)^2 \big]\partial_{\alpha^+} \nonumber \\
&&+ i \big[y^4-4y^2\beta^-\beta^+\cosh (2\gamma)+4y^2\beta^-x^+\sinh (2\gamma) \big]\partial_{\alpha^-} \nonumber \\ 
&&+i \big[-16\alpha^+\beta^++4x^+(\beta^+)^2+4(x^+)^3\big]\partial_{\beta^+} \nonumber 
\\ && +iy^2\big[-4\beta^+ \sinh (2\gamma)+4x^+\cosh (2\gamma)\big]\partial_{\beta^-}-i \big[4(\beta^+)^3-12\beta^+(x^+)^2+16x^+\alpha^+\big]\partial_{x^+} \nonumber \\
&&+iy^2 \big[-4\beta^+\cosh (2\gamma)+4x^+\sinh (2\gamma)\big]\partial_{x^-}+i\big[-6(\beta^+)^2+6(x^+)^2\big]\partial_{\gamma} \nonumber \\
&& +i\big[-8\alpha^++4x^+\beta^+\big] \, y\partial_y \label{differentialop}
\end{eqnarray}
These generators satisfy the algebra (\ref{W3wedgealgebra}). Generators $\overline{L}^h_n$, $\overline{W}^h_n$ are obtained from the above by interchanges, $x^+ \leftrightarrow x^-$, 
$\alpha^+ \leftrightarrow \alpha^-$, $\beta^+ \leftrightarrow \beta^-$.

\section{Klein-Gordon Equation for a Scalar Field in the 8d Space-time}
\hspace{5mm}
Here an  explicit form of the Klein-Gordon equation for $|\Phi\rangle$, (\ref{KGsymbolical}), is presented. 
\begin{eqnarray}
&& \Big[ y^2\partial_y^2-7y\partial_y+3\partial_{\gamma}^2-4y^2\cosh 2\gamma \, \partial_{x^+}\partial_{x^-}-4y^2\cosh 2\gamma \, (\beta^-\partial_{x^+}\partial_{\alpha^-}+\beta^+\partial_{x^-}\partial_{\alpha^+}  ) \nonumber \\
&&-4y^2\sinh 2\gamma (\partial_{x^+} \partial_{\beta^-}+\partial_{x^-} \partial_{\beta^+})
-4y^2\sinh 2\gamma (\beta^+\partial_{\alpha^+}\partial_{\beta^-}+\beta^-\partial_{\alpha^-}\partial_{\beta^+})\nonumber \\
&& +(y^4-4\beta^+\beta^-y^2 \cosh 2\gamma)\partial_{\alpha^+}\partial_{\alpha^-}-4y^2 \cosh 2\gamma \partial_{\beta^+}\partial_{\beta^-}  -m^2\Big] \, |\Phi\rangle=0  \label{KG}
\end{eqnarray}

\section{Spin-3 Field $\phi^{\mu\nu\lambda}$ in (\ref{thirdorder})}
\hspace{5mm}
An explicit form of the spin-3 field in 8d space appearing in the equation (\ref{thirdorder}) for the cubic Casimir operator  is given by\footnote{The indices of $\phi^{\mu\nu\lambda}$ are  lowered by $g_{\mu\nu}$.}
\begin{eqnarray}
\phi &\equiv & \phi_{\mu\nu\lambda} \, dx^{\mu}\, dx^{\mu} \, dx^{\lambda} \nonumber \\
&=& \frac{-1}{36y^4} \, \Big\{ -27\beta^- \, dx^- \, (dx^+)^2-27 \, \beta^+ \, dx^+(dx^-)^2+27y\, \sinh 2\gamma \, dydx^+dx^- \nonumber \\
&& +27(dx^+)^2d\alpha^-+27(dx^-)^2d\alpha^+-27d\alpha^+(d\beta^-)^2-27d\alpha^-(d\beta^+)^2 \nonumber \\
&&+72d\alpha^+d\alpha^-d\gamma+18y^2dy^2d\gamma-2y^4d\gamma^3+27\beta^-dx^-(d\beta^+)^2 +27\beta^+dx^+(d\beta^-)^2\nonumber \\
&&-72\beta^-dx^-d\alpha^+d\gamma-72\beta^+dx^+d\alpha^-d\gamma +72\beta^+\beta^-dx^+dx^-d\gamma  \nonumber \\
&&+ \Big(-27ydx^+d\beta^-dy-27ydx^-d\beta^+dy+9y^2dx^+dx^-d\gamma+9y^2d\beta^+d\beta^-d\gamma\Big) \, \cosh 2\gamma \nonumber \\
&&+\Big(27yd\beta^+d\beta^-dy-9y^2dx^+d\beta^-d\gamma-9y^2dx^-d\beta^+d\gamma\Big) \sinh2\gamma\Big\}.  \label{phiCubic}
\end{eqnarray}

\section{Equations for $\psi^{(1)a}$ and $\overline{\psi}^{(1)a}$}
\hspace{5mm}
By substituting (\ref{gau1}) and (\ref{gau2}) into (\ref{flat+}) and a similar equation for $\overline{\psi}$, explicit equations for $\psi^{(1)a}$ and $\overline{\psi}^{(1)a}$ are obtained. These are listed below. 
\begin{eqnarray}
\bullet &&d\psi^{(1)1} +4(\sinh \gamma dx^+-\cosh \gamma d\beta^+) \wedge \psi^{(1)4}+d\gamma \wedge \psi^{(1)5}=0, \nonumber \\
\bullet &&d\psi^{(1)}2+2(-\cosh \gamma dx^++\sinh \gamma d\beta^+) \wedge \psi^{(1)1}+16(\beta^+dx^+-d\alpha^+) \wedge \psi^{(1)4} \nonumber \\
&&\qquad +2(-\sinh \gamma dx^++\cosh \gamma d\beta^+) \wedge \psi^{(1)5}=0, \nonumber \\
\bullet && d\psi^{(1)3}-(\cosh \gamma dx^+-\sinh \gamma d\beta^+) \wedge \psi^{(1)2}+4(\beta^+dx^+-d\alpha^+)\wedge \psi^{(1)5}-d\gamma \wedge \psi^{(1)7} \nonumber \\&& \qquad +2(-\sinh \gamma dx^++\cosh \gamma d\beta^+)\wedge \psi^{(1)6}=0, \nonumber \\
\bullet &&d\psi^{(1)4} =0, \nonumber \\
\bullet && d\psi^{(1)5} +d\gamma \wedge \psi^1-4(\cosh \gamma dx^+-\sinh \gamma d\beta^+)\wedge \psi^{(1)4}=0, \nonumber \\
\bullet && d\psi^{(1)6} +3(-\sinh \gamma dx^++\cosh \gamma d\beta^+) \wedge \psi^{(1)1}-3(\cosh \gamma dx^+-\sinh \gamma d\beta^+)\wedge  \psi^{(1)5}=0, \nonumber \\
\bullet && d\psi^{(1)7}+4(-\beta^+ dx^++d\alpha^+) \wedge \psi^{(1)1}+(-\sinh \gamma dx^++\cosh \gamma d\beta^+)\wedge \psi^{(1)2}-d\gamma \wedge \psi^{(1)3} \nonumber \\
&& \qquad -2(\cosh \gamma dx^+-\sinh \gamma d\beta^+)\wedge \psi^{(1)6}=0, \nonumber \\ 
\bullet &&d\psi^{(1)8}+2(-\beta^+dx^++d\alpha^+) \wedge \psi^{(1)2}+(\sinh \gamma dx^+-\cosh \gamma d\beta^+) \wedge \psi^{(1)3} \nonumber \\
&&\qquad -(\cosh \gamma dx^+-\sinh \gamma d\beta^+)\wedge \psi^{(1)7}=0 \label{E1}
\end{eqnarray}
The equation for $\overline{\psi}^{(1)a}$ is obtained by the following replacement.
\begin{eqnarray}
&& x^+ \rightarrow x^-, \quad \alpha^+ \rightarrow \alpha^-, \quad \beta^+ \rightarrow \beta^-, \nonumber \\
&& \psi^{(1)1} \rightarrow \overline{\psi}^{(1)3}, \quad 
\psi^{(1)2} \rightarrow \overline{\psi}^{(1)2}, \quad 
\psi^{(1)3} \rightarrow \overline{\psi}^{(1)1}, \quad 
\psi^{(1)4} \rightarrow \overline{\psi}^{(1)8}, \nonumber \\
&& \psi^{(1)5} \rightarrow \overline{\psi}^{(1)7}, \quad 
\psi^{(1)6} \rightarrow \overline{\psi}^{(1)6}, \quad 
\psi^{(1)7} \rightarrow \overline{\psi}^{(1)5}, \quad 
\psi^{(1)8} \rightarrow \overline{\psi}^{(1)4} \nonumber 
\end{eqnarray}

\begin{eqnarray}
\psi^{(1)1} &=& \sinh \gamma \{-4Q_1 dx^++dQ_3\}+\cosh \gamma \{dQ_2+4Q_1d\beta^+\}, \nonumber \\
\psi^{(1)2} &=& (-16\beta^+ Q_1-2Q_2)dx^++16Q_1d\alpha^+-2Q_3d\beta^++dQ_4, \nonumber \\
\psi^{(1)3}& =& \cosh \gamma \{(-4\beta^+Q_3-Q_4)dx^+-2Q_5d\beta^++4Q_3d\alpha^++dQ_6\} \nonumber \\
&&+\sinh \gamma \{(-4\beta^+Q_2+2Q_5)dx^++Q_4d\beta^++4Q_2d\alpha^+-dQ_7\}, \nonumber \\
\psi^{(1)4} &=& -dQ_1, \nonumber \\
\psi^{(1)5} &=& -\sinh \gamma (4Q_1 d\beta^++dQ_2)+\cosh \gamma (4Q_1dx^+-dQ_3), \nonumber \\
\psi^{(1)6} &=& 3Q_3 dx^++3Q_2 d\beta^+-dQ_5, \nonumber \\
\psi^{(1)7} &=& \cosh \gamma \{(-4\beta^+Q_2+2Q_5)dx^++Q_4d\beta^++4Q_2d\alpha^+-dQ_7\} \nonumber \\
&&+\sinh \gamma \{(-Q_4-4\beta^+Q_3)dx^+-2Q_5d\beta^++4Q_3d\alpha^++dQ_6\}, \nonumber \\
\psi^{(1)8} &=& (-2\beta^+Q_4+Q_7)dx^++2Q_4d\alpha^+-Q_6d\beta^+-dQ_8. \label{psia}
\end{eqnarray}
Here $Q_n$ $(n=1,\ldots,8)$ are arbitrary functions. 

Similarly, general solution to (\ref{p2}) is given by
\begin{eqnarray}
\overline{\psi}^{(1)1}  &=& \cosh \gamma \{-4\beta^-\overline{Q}_3 dx^--\overline{Q}_4dx^--2\overline{Q}_5d\beta^-+4\overline{Q}_3d\alpha^-+d\overline{Q}_6\} \nonumber \\
&& +\sinh \gamma \{ -4\beta^-\overline{Q}_2dx^-+2\overline{Q}_5dx^-+\overline{Q}_4d\beta^-+4\overline{Q}_2d\alpha^--d\overline{Q}_7\}, \nonumber \\
\overline{\psi}^{(1)2} &=& -16\beta^-\overline{Q}_1dx^--2\overline{Q}_2dx^-+16\overline{Q}_1d\alpha^--2\overline{Q}_3d\beta^-+d\overline{Q}_4, \nonumber \\
\overline{\psi}^{(1)3}&=& \cosh \gamma \{4\overline{Q}_1d\beta^-+d\overline{Q}_2\}+\sinh \gamma \{-4\overline{Q}_1dx^-+d\overline{Q}_3\}, \nonumber \\
\overline{\psi}^{(1)4} &=& (-2\beta^-\overline{Q}_4+\overline{Q}_7)dx^-+2\overline{Q}_4d\alpha^--\overline{Q}_6d\beta^--d\overline{Q}_8, \nonumber \\
\overline{\psi}^{(1)5} &=&\sinh \gamma \{-4\beta^-\overline{Q}_3dx^--\overline{Q}_4dx^--2\overline{Q}_5d\beta^-+4\overline{Q}_3d\alpha^-+d\overline{Q}_6\}, \nonumber \\
&& -\cosh \gamma \{4\beta^-\overline{Q}_2dx^--2\overline{Q}_5dx^--\overline{Q}_4d\beta^--4\overline{Q}_2d\alpha^-+d\overline{Q}_7\}, \nonumber \\
\overline{\psi}^{(1)6}&=&3\overline{Q}_2d\beta^-+3\overline{Q}_3dx^--d\overline{Q}_5, \nonumber \\
\overline{\psi}^{(1)7} &=& -\sinh \gamma \{4\overline{Q}_1d\beta^-+d\overline{Q}_2\}+\cosh \gamma \{4\overline{Q}_1dx^--d\overline{Q}_3\}, \nonumber \\
\overline{\psi}^{(1)8} &=& -d\overline{Q}_1. \label{psibara}
\end{eqnarray}
Hence there are 16 perturbative modes $Q_n$, $\overline{Q}_n$ in the classical solutions. Actually, these are gauge modes. However, when  those modes are changed by amounts which are not single-valued functions on the torus, then the flat connections before and after the change are inequivalent. For static or stationary black hole solutions the functions $Q_n$ and $\overline{Q}_n$ have to be chosen such that $\psi^{(1)a}$ and $\overline{\psi}^{(1)a}$ are periodic 
in  $x^{\pm}$.

\section{Black Hole Solution without Spin-3 Charge }
\hspace{5mm}
The flat connections for a black hole without spin-3 charge are obtained by choosing suitable $Q_n$ and $\overline{Q}_n$ in (\ref{psia}) and (\ref{psibara}), which yield static or stationary connections. 
\begin{eqnarray}
Q_1=&&Q_3=0,  \nonumber \\
Q_2=&&- a \, x^+,  \nonumber \\
Q_4=&&-a(x^+)^2,   \nonumber \\
Q_5=&&-3ax^+ \, \beta^+,  \nonumber \\
Q_6=&&-\frac{1}{3}a(x^+)^3-3ax^+(\beta^+)^2,  \nonumber \\
Q_7=&&-a(x^+)^2\beta^+-4a x^+\alpha^+,  \nonumber \\
Q_8=&&\frac{1}{3} \, a(x^+)^3\beta^+-2a(x^+)^2\alpha^++ax^+(\beta^+)^3, \label{Qno-1}
\end{eqnarray}
For simplicity the variable $\gamma$ is not included in $Q_n$. 
$\overline{Q}_n$ is obtained by the following replacement.
\begin{eqnarray}
&& Q_n \rightarrow \overline{Q}_n,  \quad x^+ \rightarrow x^-, \quad \alpha^+ \rightarrow \alpha^-, \quad \beta^+ \rightarrow \beta^-, \quad a \rightarrow -\bar{a}  \nonumber 
\end{eqnarray}

The first order perturbation $\psi^{(i)a}$ for the black hole solution without spin-3 charge which are obtained by substituting (\ref{Qno-1}) into (\ref{psia}) is given as follows. 
Then $\psi^{(1)a}$ and $\overline{\psi}^{(1)a}$ are given by 
\begin{eqnarray}
\psi^{(1)1}&=& -a \cosh \gamma \, dx^+, \nonumber \\
\psi^{(1)2}&=&\psi^{(1)4}=0,  \nonumber \\
\psi^{(1)3}&=&-a\{3(\beta^+)^2 \cosh \gamma-4\alpha^+\sinh \gamma\}dx^+,  \nonumber \\
\psi^{(1)5}&=&a\sinh \gamma dx^+,   \nonumber \\
\psi^{(1)6}&=&3a\beta^+dx^+,   \nonumber \\
\psi^{(1)7}&=& a\{ 4\alpha^+\cosh \gamma-3(\beta^+)^2\sinh\gamma\}dx^+,  \nonumber \\
\psi^{(1)8}&=&-a(\beta^+)^3dx^+,
\label{psino1}
\end{eqnarray}
$\overline{\psi}^{(1)a}$ is obtained by the following replacement. 
\begin{eqnarray}
&& x^+ \rightarrow x^-, \quad \alpha^+ \rightarrow \alpha^-, \quad \beta^+ \rightarrow \beta^-, \quad a \rightarrow -\bar{a}, \nonumber \\
&& \psi^{(1)1} \rightarrow \overline{\psi}^{(1)3}, \quad 
\psi^{(1)2} \rightarrow \overline{\psi}^{(1)2}, \quad 
\psi^{(1)3} \rightarrow \overline{\psi}^{(1)1}, \quad 
\psi^{(1)4} \rightarrow \overline{\psi}^{(1)8}, \nonumber \\
&& \psi^{(1)5} \rightarrow \overline{\psi}^{(1)7}, \quad 
\psi^{(1)6} \rightarrow \overline{\psi}^{(1)6}, \quad 
\psi^{(1)7} \rightarrow \overline{\psi}^{(1)5}, \quad 
\psi^{(1)8} \rightarrow \overline{\psi}^{(1)4} \nonumber 
\end{eqnarray}
In the above equations parameters $a$ and $\bar{a}$ are the following constants. 
\begin{eqnarray} 
a &=&  2G \, (M+J), \label{aF}\\
\bar{a}  &=& 2G \, (M-J) \label{abarF}
\end{eqnarray}
Here $M$ and $J$ are mass and angular momentum, $G$ a Newton constant.

As explained in sec. 5 the perturbation series stops at the first order and the above results give exact solutions. 
The gauge connections $A$, $\overline{A}$ with $y$ components are obtained by the gauge transformations (\ref{gauge1}) and (\ref{gauge2}). The vielbein $e=\frac{1}{2}\, (A-\overline{A})$ then yields the metric $g_{\mu\nu}= (1/2) \text{tr} (e)^2$.
\begin{eqnarray}
ds^2 & = & ds^2_0 \nonumber \\
&& +a(dx^+)^2+\bar{a}(dx^-)^2-2a\beta^+d\gamma dx^+-2\bar{a}\beta^-d\gamma dx^-  \nonumber \\&&
+\frac{4}{y^4} \Big\{-a\beta^-(\beta^+)^3 -\bar{a} \beta^+(\beta^-)^3 \Big\} dx^+dx^- \nonumber \\
&& -\frac{a}{y^2} \Big \{ -3(\beta^+)^2\cosh 2\gamma +4\alpha^+ \sinh 2\gamma \Big\} dx^+dx^- \nonumber \\
&&- \frac{\bar{a}}{y^2} \Big\{ -3(\beta^-)^2\cosh 2\gamma +4\alpha^- \sinh 2\gamma \Big\} dx^+dx^- \nonumber \\
&&-\frac{a}{y^2}\Big\{ 3(\beta^+)^2\sinh 2\gamma -4\alpha^+ \cosh 2\gamma \Big \} dx^+d\beta^- \nonumber \\
&& -\frac{\bar{a}}{y^2}\Big\{ 3(\beta^-)^2\sinh 2\gamma-4\alpha^- \cosh 2\gamma \Big\} dx^-d\beta^+ \nonumber \\
&&+\frac{4a}{y^4} (\beta^+)^3d\alpha^-dx^+
+\frac{4\bar{a}}{y^4}(\beta^-)^3d\alpha^+dx^- \nonumber \\
&& -a\bar{a}\Big\{ y^2 \cosh 2\gamma-6\beta^+\beta^--\frac{4}{y^4}(\beta^+\beta^-)^3 \nonumber \\
&& \qquad +\frac{1}{y^2} \{ 9(\beta^+\beta^-)^2 \cosh 2\gamma +16\alpha^+\alpha^- \cosh 2\gamma  \nonumber \\
&& \qquad 
-12(\beta^+)^2\alpha^-\sinh 2\gamma    
-12 (\beta^-)^2\alpha^+\sinh2\gamma\} \Big\} dx^+dx^-  \label{btz}
\end{eqnarray}
where $ds^2_0$ is the metric (\ref{8dmetric}).

\section{Black Hole Solution with Spin-3 Charge up to Third Order}
\hspace{5mm}
Functions $Q_n$ and $\overline{Q}_n$ for static or stationary black holes with spin-3 charge are given as follows. It turned out it is necessary to include higher order corrections to $\psi^a$ and $\overline{\psi}^a$.  Higher-order corrections  $\psi^{(i)a}$, $\overline{\psi}^{(i)a}$ are also presented. 
\begin{eqnarray}
Q_{1}=&&b x^+,  \nonumber \\
Q_{2}=&&- ax^+-4b x^+\beta^+-8b\alpha^+,  \nonumber \\
Q_{3}=&&2b(x^+)^2,   \nonumber \\
Q_{4}=&&-a(x^+)^2+4b(x^+)^2\beta^+-16bx^+\alpha^+,   \nonumber \\
Q_{5}=&&-3ax^+ \, \beta^++2b(x^+)^3-6bx^+(\beta^+)^2-24b\alpha^+\beta^+,  \nonumber \\
Q_{6}=&&-\frac{1}{3}a(x^+)^3-3ax^+(\beta^+)^2-8b(x^+)^2\alpha^++4b(x^+)^3\beta^+-4bx^+(\beta^+)^3-24b\alpha^+(\beta^+)^2,  \nonumber \\
Q_{7}=&&-a(x^+)^2\beta^+-4a x^+\alpha^+-16bx^+\alpha^+\beta^+-16b(\alpha^+)^2+2b(x^+)^2(\beta^+)^2+b(x^+)^4,  \nonumber \\
Q_{8}=&&\frac{1}{3} \, a(x^+)^3\beta^+-2a(x^+)^2\alpha^++ax^+(\beta^+)^3+\mu x^++8b(x^+)^2\alpha^+\beta^+,\nonumber \\
&&-2b(x^+)^3(\beta^+)^2-16bx^+(\alpha^+)^2+\frac{b}{5}(x^+)^5 
+bx^+(\beta^+)^4+8b\alpha^+(\beta^+)^3, \label{Q-1}
\end{eqnarray}
$\overline{Q}_n$ is obtained by the following replacement.
\begin{eqnarray*}
Q_n \rightarrow \overline{Q}_n,  \quad x^+ \rightarrow x^-, \quad \alpha^+ \rightarrow \alpha^-, \quad \beta^+ \rightarrow \beta^-, \quad a \rightarrow -\bar{a}, \quad b \rightarrow \bar{b}, \quad \mu \rightarrow \bar{\mu}  
\end{eqnarray*}

First-order perturbations for flat connections $\psi^{(1)a}$, $\overline{\psi}^{(1)a}$ are given  as follows. 
\begin{eqnarray}
\psi^{(1)1}&=& -a \cosh \gamma \, dx^+-4b\beta^+\cosh\gamma\, dx^+-8b\cosh\gamma \, d\alpha^+, \nonumber \\
\psi^{(1)2}&=&0,  \nonumber \\
\psi^{(1)3}&=&-a\{3(\beta^+)^2 \cosh \gamma-4\alpha^+\sinh \gamma\}dx^+-4b(\beta^+)^3\cosh\gamma dx^+-24b(\beta^+)^2\cosh\gamma d\alpha^+,  \nonumber \\
\psi^{(1)4}&=& -bdx^+, \nonumber \\
\psi^{(1)5}&=&a\sinh \gamma dx^++4b\beta^+\sinh\gamma dx^++8b\sinh\gamma d\alpha^+,   \nonumber \\
\psi^{(1)6}&=&3a\beta^+dx^++6b(\beta^+)^2dx^++24b\beta^+d\alpha^+,   \nonumber \\
\psi^{(1)7}&=& a\{ 4\alpha^+\cosh \gamma-3(\beta^+)^2\sinh\gamma\}dx^+-4b(\beta^+)^3\sinh\gamma dx^+-24b(\beta^+)^2\sinh\gamma d\alpha^+,  \nonumber \\
\psi^{(1)8}&=&-a(\beta^+)^3dx^+-\mu dx^+-b(\beta^+)^4dx^+-8b(\beta^+)^3d\alpha^+,                   \label{psi1}
\end{eqnarray}
$\overline{\psi}^{(1)a}$ is obtained by the following replacement. 
\begin{eqnarray}
&& x^+ \rightarrow x^-, \quad \alpha^+ \rightarrow \alpha^-, \quad \beta^+ \rightarrow \beta^-, \quad a \rightarrow -\bar{a}, \quad b \rightarrow \bar{b}, \quad \mu \rightarrow \bar{\mu}, \nonumber \\
&& \psi^{(1)1} \rightarrow \overline{\psi}^{(1)3}, \quad 
\psi^{(1)2} \rightarrow \overline{\psi}^{(1)2}, \quad 
\psi^{(1)3} \rightarrow \overline{\psi}^{(1)1}, \quad 
\psi^{(1)4} \rightarrow \overline{\psi}^{(1)8}, \nonumber \\
&& \psi^{(1)5} \rightarrow \overline{\psi}^{(1)7}, \quad 
\psi^{(1)6} \rightarrow \overline{\psi}^{(1)6}, \quad 
\psi^{(1)7} \rightarrow \overline{\psi}^{(1)5}, \quad 
\psi^{(1)8} \rightarrow \overline{\psi}^{(1)4} \nonumber  
\end{eqnarray}

Second-order perturbations for flat connections are given  as follows. 
\begin{eqnarray}
\psi^{(2)1}&=& \zeta_2 b \mu \sinh \gamma d\beta^++\zeta_2 b \mu \cosh \gamma dx^+, \nonumber \\
\psi^{(2)2}&=&0, \nonumber \\
\psi^{(2)3}&=&-96ab(\alpha^+)^2\beta^+\cosh \gamma dx^+-32b\mu \alpha^+\sinh \gamma dx^+ +2\zeta_1a\mu \beta^+ \cosh \gamma dx^+\nonumber \\
&& +\big(-2\zeta_2 b\mu (\beta^+)^2dx^++4\zeta_2b\mu \beta^+d\alpha^+\big)\cosh \gamma-4\zeta_2b\mu \alpha^+\sinh \gamma dx^+, \nonumber \\
\psi^{(2)4}&=&0,\nonumber \\
\psi^{(2)5}&=& -\zeta_2 b \mu \sinh \gamma dx^+-\zeta_2 b \mu \cosh \gamma d\beta^+, \nonumber \\
\psi^{(2)6}&=& 48ab(\alpha^+)^2dx^+-\zeta_1a\mu dx^+,   \nonumber \\
\psi^{(2)7}&=& -96ab(\alpha^+)^2\beta^+\sinh\gamma\,dx^+-32b\mu\alpha^+\cosh\gamma dx^++2\zeta_1 a\mu \beta^+ \sinh \gamma dx^+  \nonumber \\
&&-4\zeta_2b\mu \alpha^+\cosh \gamma dx^++\big(-2\zeta_2b\mu (\beta^+)^2dx^++4\zeta_2b\mu \beta^+d\alpha^+\big)\sinh \gamma, \nonumber \\
\psi^{(2)8}&=&-48ab(\alpha^+\beta^+)^2dx^++\zeta_1a\mu (\beta^+)^2dx^++2\zeta_2b\mu (\beta^+)^2d\alpha^+-\frac{2}{3}\zeta_2 b\mu (\beta^+)^3dx^+, \nonumber \\
&&\label{psi2}
\end{eqnarray}
$\overline{\psi}^{(2)a}$ is obtained by the following replacement. 
\begin{eqnarray}
&& x^+ \rightarrow x^-, \quad \alpha^+ \rightarrow \alpha^-, \quad \beta^+ \rightarrow \beta^-, \quad a \rightarrow -\bar{a}, \quad b \rightarrow \bar{b}, \quad \mu \rightarrow \bar{\mu}, \nonumber \\
&& \psi^{(2)1} \rightarrow -\overline{\psi}^{(2)3}, \quad 
\psi^{(2)2} \rightarrow -\overline{\psi}^{(2)2}, \quad 
\psi^{(2)3} \rightarrow -\overline{\psi}^{(2)1}, \quad 
\psi^{(2)4} \rightarrow -\overline{\psi}^{(2)8}, \nonumber \\
&& \psi^{(2)5} \rightarrow -\overline{\psi}^{(2)7}, \quad 
\psi^{(2)6} \rightarrow -\overline{\psi}^{(2)6}, \quad 
\psi^{(2)7} \rightarrow -\overline{\psi}^{(2)5}, \quad 
\psi^{(2)8} \rightarrow -\overline{\psi}^{(2)4} \nonumber \\
&& \zeta_1 \rightarrow \bar{\zeta}_1, \quad \zeta_2 \rightarrow -\bar{\zeta}_2 \nonumber
\end{eqnarray}

Finally, third-order perturbations for flat connections are given  as follows. 
\begin{eqnarray}
\psi^{(3)1} &=& \cosh \gamma \, \Big(H_1+dQ^{(3)}_2\Big)-\sinh \gamma \, H_2 \nonumber \\ 
&&-4\zeta_3a^2\mu\beta^+\cosh\gamma\, dx^+ -8\zeta_3a^2\mu \cosh\gamma \, d\alpha^+, \nonumber \\
\psi^{(3)2} &=& 0, \nonumber \\
\psi^{(3)3} &=& \cosh \gamma \, H_3+\sinh \gamma \big( -H_4-4\beta^+ Q^{(3)}_2 dx^++4Q^{(3)}_2d\alpha^+\big) \nonumber \\ 
&&-4\zeta_3a^2\mu(\beta^+)^3\cosh\gamma dx^+-24\zeta_3a^2\mu(\beta^+)^2\cosh\gamma d\alpha^+, \nonumber \\
\psi^{(3)4} &=& \zeta_2 ab\mu \beta^+dx^++2\zeta_2b^2\mu (\beta^+)^2dx^++8\zeta_2b^2\mu\beta^+d\alpha^+-\zeta_3 \,\mu \, a^2dx^+, \nonumber \\
\psi^{(3)5} &=& \cosh \gamma \, H_2-\sinh \gamma \, \big(H_1+dQ^{(3)}_2 \big)+4\zeta_3a^2\mu\beta^+\sinh\gamma dx^++8\zeta_3a^2\mu\sinh\gamma d\alpha^+, \nonumber \\
\psi^{(3)6} &=& -48b^2\mu(8+\zeta_2)(\alpha^+)^2dx^+-8ab\mu\zeta_2(\beta^+)^3dx^+-8b^2\mu\zeta_2(\beta^+)^4dx^+\nonumber \\
&&-64b^2\mu\zeta_2(\beta^+)^3d\alpha^++3Q^{(3)}_2d\beta^++6\zeta_3a^2\mu(\beta^+)^2dx^++24\zeta_3a^2\mu\beta^+d\alpha^+, \nonumber \\
\psi^{(3)7} &=& \sinh \gamma \, H_3+\cosh \gamma \, \big(-H_4-4\beta^+Q_2^{(3)}dx^-+4Q^{(3)}_2d\alpha^+\big) \nonumber \\ 
&&-4\zeta_3a^2\mu(\beta^+)^3\sinh\gamma dx^+-24\zeta_3a^2\mu(\beta^+)^2\sinh\gamma d\alpha^+, \nonumber \\
\psi^{(3)8} &=& 48b^2\mu (\zeta_2+8)(\alpha^+)^2(\beta^+)^2dx^++256ab^2(\alpha^+)^4\beta^+dx^+-32ab\mu\zeta_1(\alpha^+)^2\beta^+dx^+\nonumber \\
&&+\frac{2}{3}b^2\mu\zeta_2(\beta^+)^6dx^++2b\mu^2\zeta_2(\beta^+)^2dx^++ab\mu\zeta_2(\beta^+)^5dx^++8b^2\mu\zeta_2(\beta^+)^5d\alpha^+ \nonumber \\
&&-\zeta_3a^2\mu(\beta^+)^4dx^+-8\zeta_3a^2\mu(\beta^+)^3d\alpha^+ \label{psi3}
\end{eqnarray}
Here one-forms $H_i$ $(i=1,2,3,4$) and the function $Q^{(3)}_2$ are as follows.
\begin{eqnarray}
H_1 &=& 5ab\mu \zeta_2 (\beta^+)^2 dx^++\frac{20}{3}b^2\mu\zeta_2(\beta^+)^3dx^++40b^2\mu\zeta_2(\beta^+)^2d\alpha^+, \nonumber \\
H_2 &=& 256ab^2(\alpha^+)^3dx^+-16ab\mu\zeta_1\alpha^+dx^+, \nonumber \\
H_3 &=&96b^2\mu(\zeta_2+8)(\alpha^+)^2\beta^+dx^++256ab^2(\alpha^+)^4dx^+-32ab\mu\zeta_1(\alpha^+)^2dx^+\nonumber \\
&&+4b^2\mu\zeta_2(\beta^+)^5dx^++5ab\mu\zeta_2(\beta^+)^4dx^++4b\mu^2\zeta_2\beta^+dx^++40b^2\mu\zeta_2(\beta^+)^4d\alpha^+, \nonumber \\
H_4&=& -256ab^2(\alpha^+)^3(\beta^+)^2dx^++4ab\mu(4\zeta_1+\zeta_2)\alpha^+(\beta^+)^2dx^+, \nonumber \\
Q^{(3)}_2&=&-256ab^2(\alpha^+)^3\beta^++4ab\mu(4\zeta_1+\zeta_2)\alpha^+\beta^+
\end{eqnarray}
$\overline{\psi}^{(3)a}$ is obtained by the following replacement of the above. 
\begin{eqnarray}
&& x^+ \rightarrow x^-, \quad \alpha^+ \rightarrow \alpha^-, \quad \beta^+ \rightarrow \beta^-, \quad a \rightarrow -\bar{a}, \quad b \rightarrow \bar{b}, \quad \mu \rightarrow \bar{\mu}, \nonumber \\
&& \psi^{(3)1} \rightarrow \overline{\psi}^{(3)3}, \quad 
\psi^{(3)2} \rightarrow \overline{\psi}^{(3)2}, \quad 
\psi^{(3)3} \rightarrow \overline{\psi}^{(3)1}, \quad 
\psi^{(3)4} \rightarrow \overline{\psi}^{(3)8}, \nonumber \\
&& \psi^{(3)5} \rightarrow \overline{\psi}^{(3)7}, \quad 
\psi^{(3)6} \rightarrow \overline{\psi}^{(3)6}, \quad 
\psi^{(3)7} \rightarrow \overline{\psi}^{(3)5}, \quad 
\psi^{(3)8} \rightarrow \overline{\psi}^{(3)4} \nonumber \\
&& \zeta_1 \rightarrow \bar{\zeta}_1, \quad \zeta_2 \rightarrow -\bar{\zeta}_2, \quad
\zeta_3 \rightarrow \bar{\zeta}_3 \nonumber
\end{eqnarray}
$\zeta_i$ and $\overline{\zeta}_i$ are constants to be determined at (\ref{zetas}) and (\ref{zetabars}) in the main text by the integrability conditions for the partition function.

\section{Metric for Black Hole Solution with Spin-3 Charge up to First-Order Perturbation}
\hspace{5mm}
Up to the first order of perturbation the metric of the black hole solution with spin-3 charge is given by 
\begin{equation}
ds^2 = ds_0^2+ds_1^2, \label{as}
\end{equation}
where $ds^2_0$ is given in (\ref{8dmetric}) and the other term is given as follows.
\begin{eqnarray}
ds_1^2&=&a(dx^+)^2+\bar{a}(dx^-)^2  \nonumber \\
&+&\Big[4\frac{-a(\beta^+)^3\beta^--\bar{a}\beta^+(\beta^-)^3-b(\beta^+)^4\beta^-+\bar{b}\beta^+(\beta^-)^4+\beta^+\bar{\mu}-\beta^-\mu}{y^4}  \nonumber \\
&& \ +\frac{3a(\beta^+)^2+3\bar{a}(\beta^-)^2 +4by^2(\beta^+)^3-4\bar{b}y^2(\beta^-)^3   }{y^2}\cosh\gamma \nonumber \\
&&  \qquad  -\frac{4a\alpha^++4\bar{a}\alpha^-}{y^2}\sinh\gamma \Big]dx^+dx^-  \nonumber\\
&+&12bdx^+d\alpha^+-12\bar{b}dx^-d\alpha^- \nonumber \\
&+&2\beta^+\Big(-a-2b\beta^+\Big)dx^+d\gamma -2\beta^-\Big(\bar{a}-2\bar{b}\beta^-\Big)dx^-d\gamma  \nonumber \\
&+&\frac{4\left[a(\beta^+)^3+(\beta^+)^4 b+2(\beta^-)^2 \bar{b} \left(4 \beta^+ \beta^--3 y^2\cosh 2\gamma\right)+\mu\right]}{y^4}dx^+d\alpha^-\nonumber \\
&+&\frac{\left[4 a\alpha^+ \cosh2\gamma-(\beta^+)^2 (3a+4b\beta^+)\sinh2\gamma\right]}{y^2}dx^+d\beta^-    \nonumber \\
&+&\frac{4\left[\bar{a}(\beta^-)^3-(\beta^-)^4 \bar{b}-2(\beta^+)^2 b \left(4 \beta^- \beta^+-3 y^2\cosh 2\gamma\right)-\bar{\mu}\right]}{y^4}dx^-d\alpha^+\nonumber \\ &+&\frac{\left[4 \bar{a} \alpha^- \cosh2\gamma-(\beta^-)^2 (3\bar{a} -4\bar{b}\beta^-)\sinh2\gamma\right]}{y^2}dx^-d\beta^+  \nonumber \\
&+&\frac{32 \left[(\beta^+)^3 b-(\beta^-)^3 \bar{b}\right]}{y^4}d\alpha^+d\alpha^--16b\beta^+d\alpha^+d\gamma+16\bar{b}\beta^-d\alpha^-d\gamma \nonumber \\
&-& \frac{24}{y^2}\, b(\beta^+)^2 \, \sinh 2\gamma d\alpha^+d\beta^-+\frac{24}{y^2}\, \bar{b}(\beta^-)^2 \, \sinh 2\gamma d\alpha^-d\beta^+,
\end{eqnarray}

Parameters $a$, $\bar{a}$ are related to the mass and the angular momentum as in (\ref{a}), (\ref{bara}), and  $b$ and $\bar{b}$  to the spin-3 charges. $\mu$ and $\bar\mu$ are chemical potentials for the charges. The result for the spin-3 gauge field is not presented. It is also checked that $\varphi$ satisfies the equation, $\nabla_{\mu} \, \varphi_{\nu\lambda\rho}=0$.

\end{document}